\documentclass[12pt,preprint]{aastex}
\usepackage{natbib}
\newcommand{\eqb}{\begin{equation}}
\newcommand{\eqe}{\end{equation}}

\shorttitle{Hydraulic/Shock Jumps}
\shortauthors{Boley and Durisen}

\begin{document}
\title{Hydraulic/Shock Jumps in Protoplanetary Disks}
\author{A.\ C.\ Boley and R.\ H.\ Durisen}
\affil{Astronomy Department, Indiana University, 727 East Third Street, Bloomington, IN 47405-7105}
\slugcomment{Accepted for publication in ApJ}

\begin{abstract}

In this paper, we describe the nonlinear outcome of spiral shocks in protoplanetary disks.  Spiral shocks, for most protoplanetary disk conditions, create a loss of vertical force balance in the post-shock region and result in rapid expansion of the gas perpendicular to the disk midplane.  This expansion has characteristics similar to hydraulic jumps, which occur in incompressible fluids.  We present a theory to describe the behavior of these hybrids between shocks and hydraulic jumps (shock bores) and then compare the theory to three-dimensional hydrodynamics simulations.  We discuss the fully three-dimensional shock structures that shock bores produce and discuss possible consequences for disk mixing, turbulence, and evolution of solids.

\end{abstract}

\keywords{accretion, accretion disks --- convection --- hydrodynamics --- shock waves --- solar system: formation --- turbulence}

\section{INTRODUCTION\label{intro}}

To address the dynamic evolution of protoplanetary disks properly, the three-dimensionality of spiral waves and shocks must be understood.
Fully three-dimensional waves in accretion disks have been studied by several authors \citep[e.g.,][]{lubow1981,linetal1990waves,korycanskyandpringle1995,lubowogilvie1998,ogilvie2002331,ogilvie2002330,
bate2002waves}, typically in the context of tidal forcing.   It has been noted \citep[see][]{lubowogilvie1998} that these waves act like fundamental modes ($f$-modes), which correspond to large surface distortions in the disk.  These waves can affect the disk's evolution through wave dissipation at the disk's surface and through gap formation \citep{bryden1999,bate2003}.  
The ability of radially propagating waves to transport angular momentum, and also influence gap formation, is dependent on the thermal stratification of the disk \citep[e.g.][]{linetal1990waves,lubowogilvie1998}.  Thus, the vertical direction, although often ignored, plays a very important role. 

Independent work, in the context of gravitational instabilities (GIs), has also found that, once GIs become well delveloped, e.g., in the {\it asymptotic state} of \citet{mejia2005}, the spiral waves resulting from GIs in the disk also behave like f-modes and involve large surface distortions \citep{pickett1998,pickett2000,pickett2003,durisengireview}.  Furthermore, it was noted in these studies that the spiral wave activity is highly nonlinear and involves many modes of comparable strengths \citep[see also][]{gammie2001,lodato_rice2004}.  In the asymptotic disk, sudden increases in disk scale height occur. These splashes of disk material seem to be related to shocks in the disk and could have important consequences for disk evolution through the generation of turbulence, breaking waves, and chondrule formation \citep{boley2005proc}.  In order to understand this shock-related splashing, which is evidently highly nonlinear, an approach other than $f$-mode analysis is required.

\citet[hereafter MC98]{martos1998} investigated shocks in the Galactic disk, where the otherwise isothermal equation of state (EOS) is stiffened by magnetic fields.  Their findings indicate that shocks occurring in semi-compressible fluid disks have characteristics of hydraulic jumps and behave, in part, like gravity modes ($g$-modes).   A classical hydraulic jump occurs in an incompressible fluid, in which the only way to reduce the kinetic energy of fluid elements coming into the wave is to convert it to gravitational potential or turbulent energy \citep{massey1970}.  Examples of hydraulic jumps can be found in spillways when there is an abrupt change in slope, and the slowly moving water has a greater height than the rapidly moving water.  The hydraulic jump is the abrupt change in the height of the flowing water.  The same phenomenon can be exhibited in a disk because, like the flowing water, the unperturbed disk is usually hydrostatic in the $z$-direction and  work must be done against gravity to expand the disk in the vertical direction.  Abrupt changes in the scale height of a disk are thus possible for similar reasons.  MC98 present an analytical theory, along with two-dimensional magnetohydrodynamics simulations, in the context of hydraulic jump conditions, and note that the jumping gas can lead to high-altitude shocks when the jumping material crashes back onto the disk.  After MC98, \citet{gomezcox2002} simulated a portion of the Galactic disk in three dimensions, noting a behavior of the gas similar to that found by MC98.  However, their analyses neglect self-gravity, which, as we show, can affect the morphology of a global shock. 

In this paper we present an analysis of hydraulic/shock jumps, which we call {\it shock bores}, in protoplanetary disks.  The theory laid out here combines the Rankine-Hugoniot shock and hydraulic jump conditions for both the adiabatic and isothermal cases.  Fully three-dimensional hydrodynamics simulations are used in \S \ref{results} to illustrate simple cases of spiral shocks in protoplanetary disks. We discuss some implications of theses results in \S \ref{discussion}. Section \ref{conclusions} summarizes our main conclusions.

\section{SHOCK BORES\label{hstheory}}
\subsection{Plane-Parallel Approximation\label{planeparallel}}
We lay out the general theory for disk bores in this section.  In a disk, the gas is semi-compressible and abrupt vertical expansions will typically occur after a shock, where the pressure gradient  in the $z$-direction is the largest.
The reason for a shock bore can be understood by considering hydrostatic equilibrium (HE) and the EOS.  For simplicity, assume that (1) the shock is planar, (2) it is propagating in the $x$-direction, (3) the disk is vertically stratified in a direction vertical $z$-perpendicular to the $x$-direction, with the pre-shock region in vertical HE, and (4), except for the discontinuity at the shock front, we ignore variations in the $x$-direction.  With conditions 3 and 4 we may write
\eqb \frac{1}{\rho}\frac{\mathrm{d}P}{\mathrm{d}z}=-\frac{\mathrm{d}\Phi}{\mathrm{d}z}\rm \label{he}
\eqe
for the pre-shock flow, where $\Phi$ is the total gravitational potential, $P$ is the pressure, and $\rho$ is the density.  Consider first the case of an adiabatic shock.  Then, we have, for the Rankine-Hugoniot shock conditions, 
\eqb \frac{P_2}{P_1}=\frac{2\gamma\mathcal{M}^2}{\gamma+1}-\frac{\gamma-1}{\gamma+1}\label{pfull}\eqe
and
\eqb \frac{u_2}{u_1}=\frac{\rho_1}{\rho_2}=\frac{\gamma-1}{\gamma+1}+\frac{2}{\gamma+1}\frac{1}{\mathcal{M}^2},\label{rhofull}\eqe
where subscripts 1 and 2 represent the pre- and post-shock regions, respectively, and where the gas speed relative to the shock is $u$, the ratio of specific heats is $\gamma$, and $\cal{M}$ is the Mach number given by
\eqb \mathcal{M}^2=\frac{\rho_1 u_1^2}{\gamma P_1}.\label{mach}\eqe  
If the pre-shock gas is initially in vertical HE, what is the state of vertical force balance behind the shock?  For the adiabatic case, using equations (\ref{pfull}) and (\ref{rhofull}), we can write the ratio of the post-shock and pre-shock pressure body forces as
\begin{eqnarray}\frac{1}{\rho_2}\frac{\mathrm{d}P_2}{\mathrm{d}z}\left(\frac{1}{\rho_1}\frac{\mathrm{d}P_1}{\mathrm{d}z}\right)^{-1} & = & \left(\frac{2\gamma\mathcal{M}^2}{\gamma+1}-\frac{\gamma-1}{\gamma+1}\right)
\left[\frac{\gamma-1}{\gamma+1}+\frac{2}{\left(\gamma+1\right)\mathcal{M}^2}\right]\\
& =& \frac{2\gamma\mathcal{M}^4\left(\gamma-1\right)-\mathcal{M}^2\left(1-6\gamma+\gamma^2\right)-2\left(\gamma-1\right)}{\mathcal{M}^2\left(\gamma+1\right)^2}.\label{pressurefraction}\end{eqnarray}
For the gravitational body force, let $\Phi_{\ast}$ be the background potential, presumably due to the central star, and let $\Phi_{g_1}$ and $\Phi_{g_2}$ be the contribution to $\Phi$ from the gas self-gravity in the pre- and postshock regions.  The ratio of the potential gradients then becomes
\begin{eqnarray}\frac{\mathrm{d}\Phi_2}{\mathrm{d}z}\left(\frac{\mathrm{d}\Phi_1}{\mathrm{d}z}\right)^{-1}
& =&  \frac{\mathrm{d}_z\Phi_{\ast}+\mathrm{d}_z\Phi_{g_2}}{
\mathrm{d}_z\Phi_{\ast}+\mathrm{d}_z\Phi_{g_1}}\\
& = & \frac{q+\mathrm{d}_z\Phi_{g_2}\left(\mathrm{d}_z\Phi_{g_1}\right)^{-1}}{q+1},\label{potalmost}
\end{eqnarray}
where
\begin{eqnarray} q &=& \frac{\mathrm{d}\Phi_*}{\mathrm{d}z}\left(\frac{\mathrm{d}\Phi_{g_1}}{\mathrm{d}z}\right)^{-1}\label{qratio}\end{eqnarray}
 is the ratio of the background  and the pre-shock gas potential gradients.  Condition 4  permits us to write $\nabla^2\Phi\approx\nabla_z^2\Phi\propto\rho$; therefore, the ratio of the self-gravity potential gradients is equivalent to the ratio of the densities.  This yields
\begin{eqnarray}\frac{\mathrm{d}\Phi_2}{\mathrm{d}z}\left(\frac{\mathrm{d}\Phi_1}{\mathrm{d}z}\right)^{-1}
 = \frac{q+a}{q+1},\label{potfraction}
\end{eqnarray}
where
\begin{eqnarray}a=\frac{\left(\gamma+1\right)\mathcal{M}^2}{2+\left(\gamma-1\right)\mathcal{M}^2}.\label{adef}\end{eqnarray}

\begin{figure}
\begin{center}
\plottwo{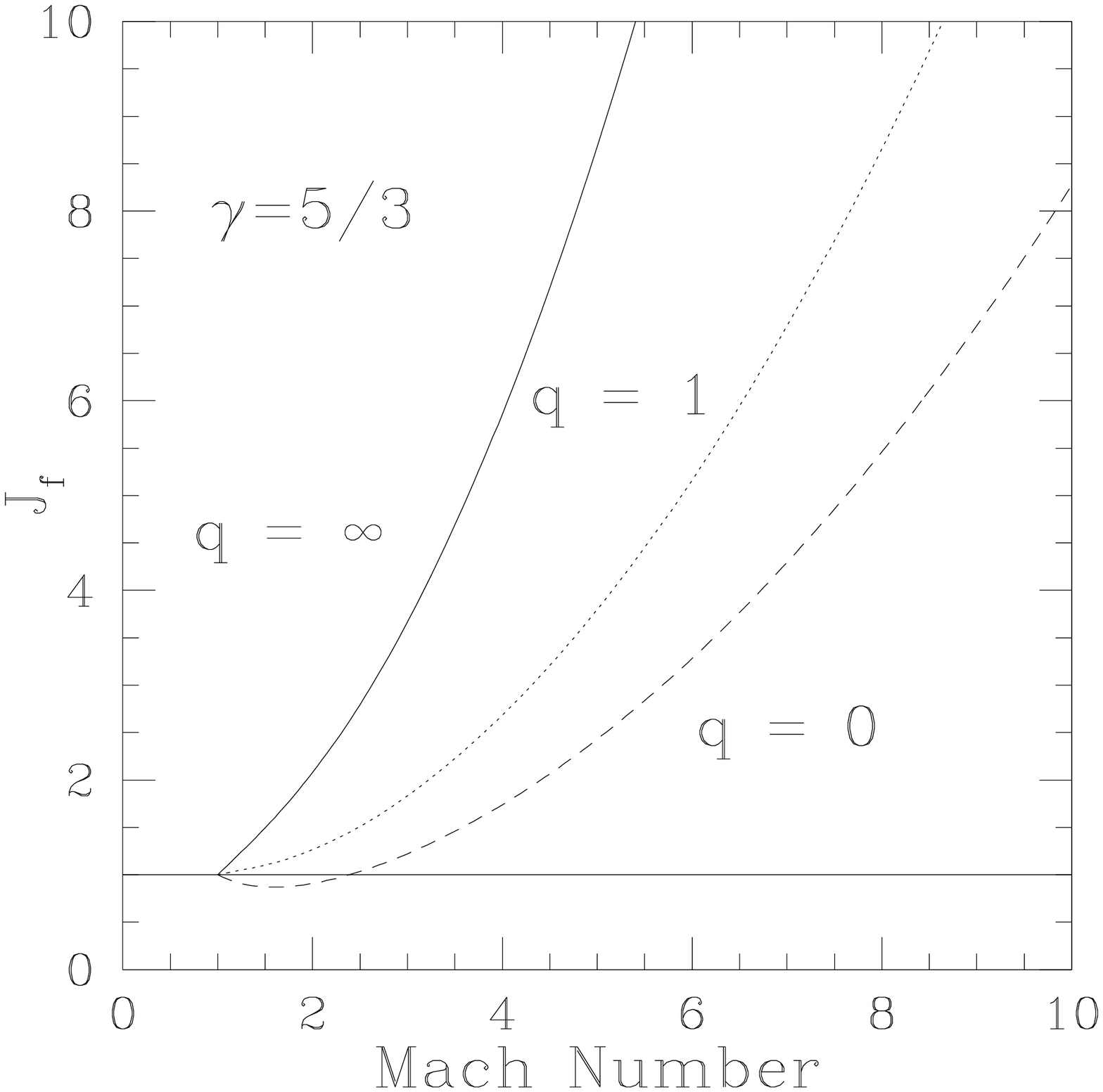}{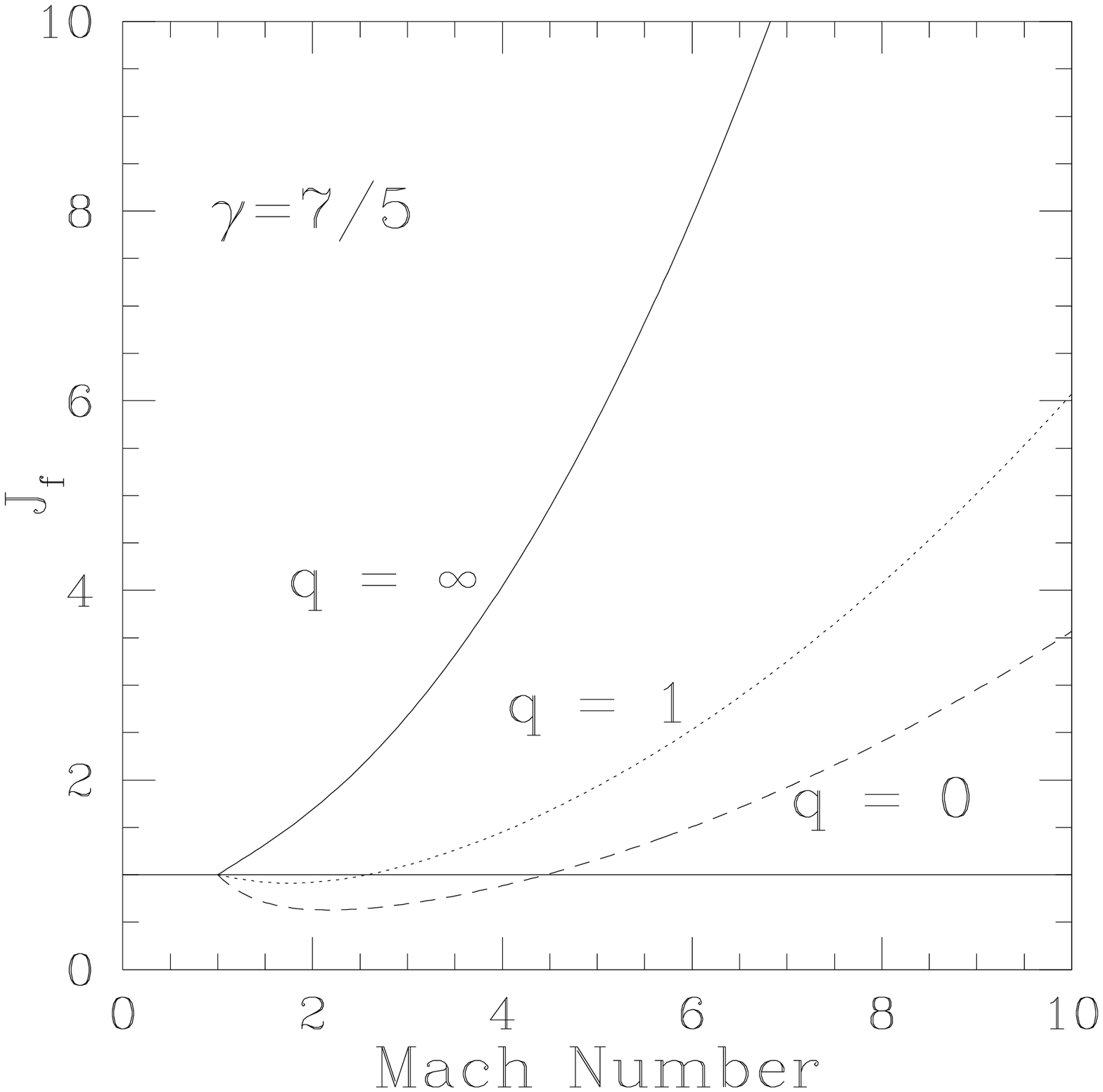}
\caption{Three curves showing the adiabatic $J_f$ with respect to $\mathcal{M}$ for $q\rightarrow\infty\rm\ (solid), \ 1\ (dotted),\ and\ 0\ (dashed)$, corresponding to no self-gravity, equal background and self-gravity, and fully self-gravitating, respectively.  Notice that for $\gamma=5/3$ (left), the disk only collapses for the self-gravity dominated case and only at low Mach numbers.  However, when $\gamma=7/5 $ (right), the compressions are noticeable at low $\mathcal{M}$ when self-gravity is comparable to the background potential. \label{fig1}}
\end{center}
\end{figure}

Define the jump-factor $J_f$ to be the ratio of equations (\ref{pressurefraction}) and (\ref{potfraction}):
\begin{eqnarray}J_f & \equiv &\frac{1}{\rho_2}\frac{\mathrm{d}P_2}{\mathrm{d}z}\left(\frac{1}{\rho_1}\frac{\mathrm{d}P_1}{\mathrm{d}z}\right)^{-1}\left[\frac{\mathrm{d}\Phi_1}{\mathrm{d}z}\left(\frac{\mathrm{d}\Phi_2}{\mathrm{d}z}\right)^{-1}\right]\label{jf1}\\
& = & \frac{2\gamma\mathcal{M}^4\left(\gamma-1\right)-\mathcal{M}^2\left(1-6\gamma+\gamma^2\right)-2\left(\gamma-1\right)}{\mathcal{M}^2\left(\gamma+1\right)^2}\left[\frac{q+1}{q+a}\right]\label{jf2}.\end{eqnarray}
Note the limits of equation (\ref{jf2}).  When self-gravity dominates ($q\rightarrow 0$) and $\mathcal{M}$ is large,
\begin{eqnarray}J_f \rightarrow \frac{2 \gamma \mathcal{M}^2\left(\gamma-1\right)^2 }
{\left(\gamma+1\right)^3},\label{jumplimit1}\end{eqnarray}
and, when the background potential dominates $(q\rightarrow\infty)$ and $\mathcal{M}$ is large,
\begin{eqnarray}J_f \rightarrow \frac{2 \gamma \mathcal{M}^2\left(\gamma-1\right) }
{\left(\gamma+1\right)^2}.\label{jumplimit2}\end{eqnarray}
To understand the significance of $J_f$, consider the vertical acceleration of the gas in the post-shock region:
\begin{eqnarray}a_z=\frac{\mathrm{d}v_z}{\mathrm{d}t}&=&-\left(\frac{1}{\rho_2}\frac{\mathrm{d}P_2}{\mathrm{d}z}+\frac{\mathrm{d}\Phi_2}{\mathrm{d}z}\right).\label{accelpart1}
\end{eqnarray}
Using equation (\ref{jf1}) in equation (\ref{accelpart1}) yields
\begin{eqnarray}a_z=\frac{\mathrm{d}v_z}{\mathrm{d}t}&=&-
\frac{\mathrm{d}\Phi_{2}}{\mathrm{d}z} \left[J_f \frac{1}{\rho_1}\frac{\mathrm{d}P_1}{\mathrm{d}z} \left( \frac{\mathrm{d}\Phi_{1}}{\mathrm{d}z} \right)^{-1}+1 \right]\\
& = &\frac{\mathrm{d}\Phi_{2}}{\mathrm{d}z}\left(J_f-1\right).\label{accelpart2}\end{eqnarray}
In the limit of no self-gravity, where $q\rightarrow\infty$, $a_z\approx\Omega^2 z \left(J_f-1\right)$ for a thin disk, where $\Omega$ is the Keplerian circular angular speed.
Equation (\ref{accelpart2}) demonstrates the physical meaning of $J_f$.  When $J_f > 1 $, the gas is overpressured and it will expand vertically, while  for $J_f < 1$, self-gravity causes the gas to compress. 
If the shock is truly strong, an expansion will always occur in the adiabatic case ($\gamma>1$).  However, self-gravity and $\gamma$ alter the strength of the adiabatic vertical acceleration (see Fig.\ \ref{fig1}).   For low $\gamma$ and $q$, regimes with  $J_f < 1$ exist at low $\mathcal{M}$.

Now consider the case of an isothermal shock, which reduces the jump-shock conditions to 
\begin{equation} \frac{P_2}{P_1}=\frac{\rho_2}{\rho_1}=\mathcal{M}^2.\label{isofull}\end{equation}
For these conditions with the same assumptions as stated for the adiabatic case, the resulting $J_f$ is
\begin{equation}J_f=\frac{q+1}{q+\mathcal{M}^2}.\label{jfiso}\end{equation}
What one can immediately see from equation (\ref{jfiso}) is that, if self-gravity is important (typically $q\lesssim 100$, which corresponds to a massive disk), one expects an isothermal gas to {\it compress} vertically in the post-shock region because $\mathcal{M} \ge 1$, but, if $q\rightarrow\infty$, spiral shocks in an isothermal disk will be essentially two-dimensional in the sense that vertical HE is maintained and the waves propagate as spiral density waves without any change in the disk scale height.   It should therefore be noted that spiral waves in protoplanetary disks will only behave like pure density waves when the self-gravity of the disk is negligible and the EOS is nearly isothermal.

To predict the height of a shock bore, consider a classical hydraulic jump.  The height of a classical hydraulic jump can be predicted using the Froude number $\rm F$ of the pre-jump region, which measures whether the flow is {\it rapid}, $\rm F>1$, or {\it tranquil}, ${\rm F} < 1$ \citep[e.g.,][]{massey1970}.  Assuming that the jump is non-dissipative, the height of the post-jump flow $h_2$ can be found by
\begin{equation} \frac{h_2}{h_1} = -\frac{1}{2}+\sqrt{\frac{1}{4}+2\rm F^2}\label{classicjump},\end{equation}
where $h_1$ is the height of the fluid in the pre-jump region.  In the limit that ${\rm F} \gg 1$, $h_2/h_1\sim \rm F$.  This classical jump result can be used as a model for understanding the maximum height a shock bore reaches during the post-shock vertical expansion. First, consider the definition of the Froude number,
\begin{equation}{\rm F} \equiv \frac{u_1}{\sqrt{g h_1}},\label{froudedef}\end{equation}
where $g$ is the acceleration of gravity and $u_1$ represents the pre-jump flow in the frame of the jump.  Second, consider a non-self-gravitating disk; in this limit, we may write $g\approx-\Omega^2z$ and $h_1\approx c_s/ \Omega$, where $\Omega$ is the circular speed of the gas, $c_s$ is the midplane sound speed, and $h_1$ is the scale height of the disk in the pre-shock region. Using these relations in place of the corresponding terms in equation (\ref{froudedef}) reveals that, for a non-self-gravitating disk, ${\rm F}\rightarrow\mathcal{M}$.   Any relation that we derive for the ratio of the scale heights before and after the expansion should have a behavior similar to equation (\ref{classicjump}), since $\rm F$ and $\mathcal{M}$ are closely related.

To relate the scale heights before and after a shock bore, consider equation (\ref{accelpart2}) to be a measure of the overpressurization in the post-shock gas.  From this perspective, $a_z$ behaves as an ``anti-gravity'' term and represents a force field capable of doing work on the gas.  Furthermore, assume that the difference in the gravitational potentials before the expansion and at the peak of the expansion also measures work done in the expansion.  Such an approximation allows us to write
\begin{eqnarray} \int_0^{h_1} \mathrm{d} z~a_z  &\approx& \int_0^{h_2} \mathrm{d} z~\Omega^2 z -
\int_0^{h_1} \mathrm{d} z~\Omega^2 z \mathrm{,}\label{scaleapprox}\\
\frac{h_2}{h_1} &\approx &\sqrt{J_f},\label{scaleheights}\end{eqnarray}
where the right-hand side of equation (\ref{scaleapprox}) is the potential difference of the gas before and after expansion and the left-hand side of equation (\ref{scaleapprox}) is a measure of the work that can be done by the post-shock overpressure. Note that we neglect self-gravity in this approximation because for many astrophysical situations $q\gtrsim100$.  This relation has a behavior similar to equation (\ref{classicjump}); when $\mathcal{M}=1$, $h_2/h_1=1$, and, when $\mathcal{M}\gg1$, $h_2/h_1\sim\mathcal{M}$.  This does not mean that every fluid element is expected to jump to the new height given by equation (\ref{scaleheights}).  Instead, the scale height, as a characteristic height of the disk, will be changed.  For example, in equation (\ref{accelpart2}), material near the midplane, where $\partial \Phi_2/\partial z\rightarrow0$, will hardly be affected by a single jump, while high altitude gas will have the strongest response.  

Equation (\ref{scaleheights}) can also be derived using mass and momentum flux arguments (see MC98).  Mass conservation requires that $\Sigma_1 u_1 = \Sigma_2 u_2$, and momentum conservation requires that $p_1-p_2$=$\Sigma_1 u_1 (u_2-u_1)$, for $p_1=\int P_1~ dz$.  Again, assume that self-gravity is negligible and that the structure before and after the shock is homologous when in equilibrium; then $P=A\rho_{0}h^2$, $p = AB \rho_{0} h^3$, and $h=C\Sigma/\rho_{0}$, where $A$, $B$, and $C$ are shape factors that depend on the details of the model and $\rho_{0}$ is the midplane density.  By using mass and momentum conservation, one can show
\begin{eqnarray}\left( \frac{h_2}{h_1} \right)^2 = \frac{u_2}{u_1}+\frac{\gamma\mathcal{M}^2}{BC}
\frac{u_2}{u_1}\left( 1-\frac{u_2}{u_1}\right).\label{scaleheights2}\end{eqnarray}
Using equation (\ref{rhofull}) and assuming $BC\approx1$, equation (\ref{scaleheights2}) becomes equation (\ref{scaleheights}).  Depending on the EOS, the assumption that $BC\approx 1$ may a bit inaccurate, but it should introduce an error in equation (\ref{scaleheights}) no larger than about 10\%.  For example, in an isentropic gas $BC =\Gamma(1+m)\Gamma(1/2 + m)/[\Gamma(3/2 + m)\Gamma(m)]^{-1}$, where $m=\gamma/(\gamma-1)$, which yields $BC=0.833$ for $\gamma=5/3$, and $BC\rightarrow 1$, as $\gamma\rightarrow1$.

\subsection{A Shock Bore in a Disk\label{realcase}}

 In a disk, shock bores turn into large waves that break onto the disk's surface, resulting in flow that is considerably more complex than that described in \S \ref{planeparallel}.  Shock bores will not only create vertical undulations in the disk, but will drive fluid elements to large radial excursions from their circular orbits and result in stirring the nebula, possibly mixing it.  For reasons we explain below, inside the corotation radius of the spiral shock, these waves flow back over the spiral shock and break onto the pre-shock flow.  This behavior is confirmed in the numerical simulations of \S\S \ref{numericaltechniques} and \ref{results}. We conjecture that shock bores will also result in breaking waves that crash onto the pre-shock flow outside the corotation radius, but we do not compute simulations outside corotation in this paper. 
 
\begin{figure}
\begin{center}
\plotone{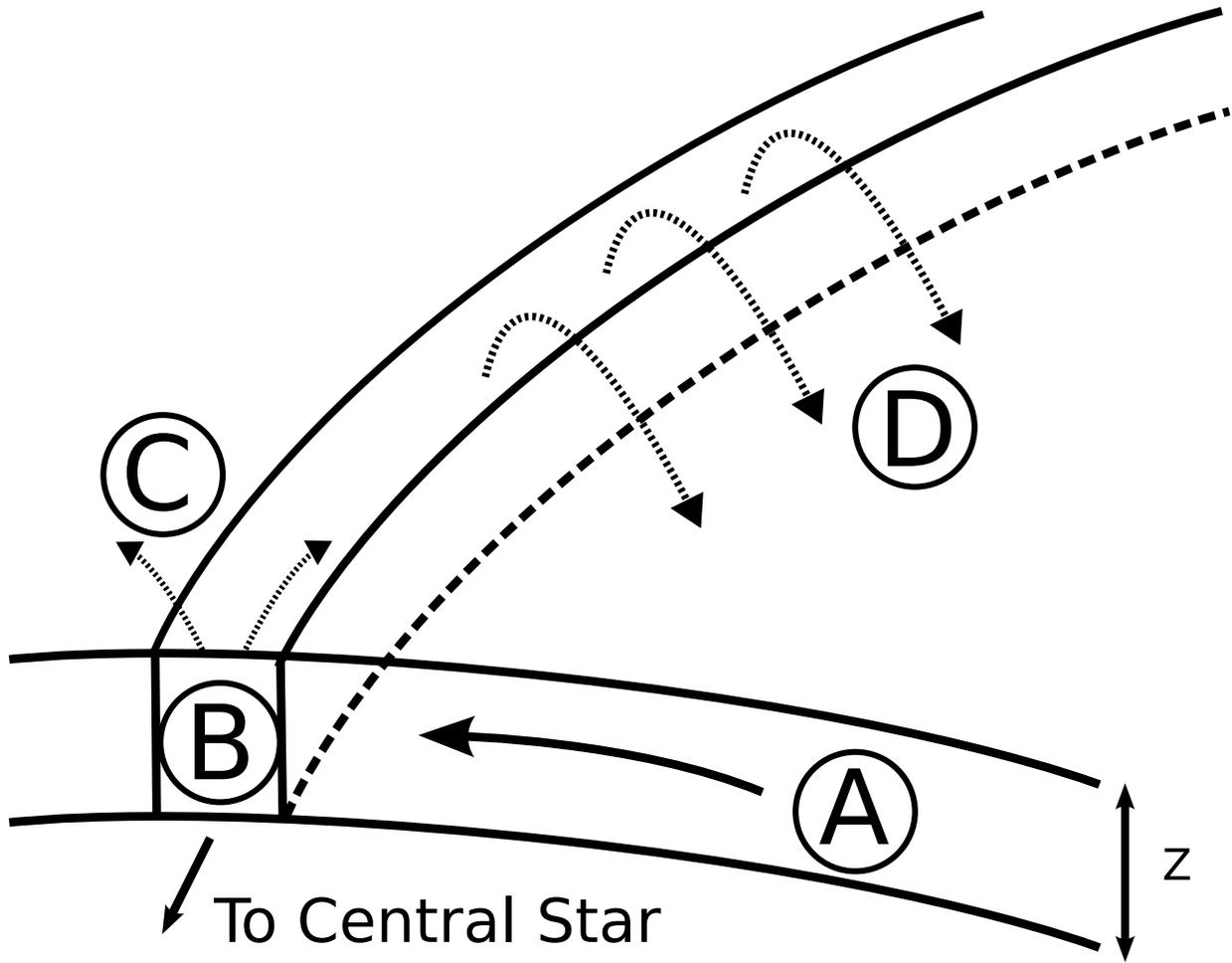}
\caption{Cartoon depicting the gas flow in an shock bore in the frame of the spiral shock inside corotation.  The gas in the pre-shock region flows into the spiral shock (A).  The shock (B) causes the material to be out of vertical force balance and a rapid expansion results (C).  Due to spiral streaming and the loss of pressure confinement, some of the gas will flow back over the spiral wave and break onto the disk in the pre-shock region at a radius inward from where it originated (D).  \label{fig2}}
\end{center}
\end{figure}

 For the following discussion, consider the point of view from inside the corotation radius of the spiral shock.  The development of breaking waves can be understood by recognizing two effects: When the gas crosses the shock front, the shock-normal component of a fluid element's velocity will be diminished by a factor given by the inverse of equation (\ref{rhofull}), while the tangential component will be preserved, allowing the flow to be supersonic after the shock.  This leads to streaming along the spiral arms, as demonstrated in streamline simulations of fluid elements in spiral galaxies \citep{robertsetal1979}.   In addition, when the gas expands upward, the pressure confinement in the direction normal to the shock front is lost, and the material expands horizontally, causing some gas to flow back over the top of the shock.  As the jumping gas moves inward and out over the pre-shock flow, it no longer has pressure support from underneath and breaks back onto the disk.  The resulting morphology is a spiral pattern moving through the disk with breaking surface waves propagating along the disk's surface with the same pattern speed as the spiral wave (see Fig.\ \ref{fig2}).   This morphology is only expected for a simple shock bore, i.e., there are no additional waves and shocks except for what are produced by cleanly defined spiral and breaking waves.  In a real disk with competing spiral waves, the behavior can be much more complex.

For such a shock bore, there should also be a radius at which the hydraulic jumps provide the strongest corrugation in the disk's surface.  Consider the fluid elements in the disk to be essentially on Keplerian orbits.  In a simple approximation, a shock bore produces a vertical perturbation to the fluid element's orbit, putting it on an inclined orbit leaving and then returning to the midplane in about half a revolution.  In the inner disk, the orbit period of a fluid element is much shorter than the pattern period of a spiral wave.  As a result, after a fluid element encounters the first spiral shock, the pattern will have moved only a small fraction of a pattern period by the time the fluid element encounters another arm; all fluid elements end up elevated between shock passages.   As one moves radially outward in the disk, the advancement of the spiral shock becomes important and shock bores develop into breaking waves.  However, as one moves toward the corotation radius, the shocks become weak, and shock bores are suppressed.  

\section{NUMERICAL TECHNIQUES}\label{numericaltechniques}
\subsection{The IUHG Code}
Our simulations are calculated using the Indiana University Hydrodynamics Group (IUHG) code  \citep[see][]{pickett1998,pickett2000,pickett2003,mejiaphd,mejia2005}.  It solves the hydrodynamics equations of motion, with self-gravity, on an Eulerian cylindrical grid $(r\mathrm{,\ } \varphi \mathrm{,\ } z)$ and is second order in space and time. Outflow boundaries are used at the inner, outer, and upper grid boundaries, and reflection symmetry is assumed about the equatorial plane.  The self-gravity component of the potential is calculated directly from the Poisson equation and a multipole expansion of spherical harmonics is used for the boundary potential with $\ell\mathrm{,}\ m\le10$ \citep{pickettphd}.   Shocks are treated in the disk using artificial bulk viscosity \citep{normanwinkler}.  The grid resolutions for the calculations presented here are (256, 128, 64), (256, 512, 32), and (256, 512, 64).  As discussed below, we scale the disks to $r = 6$ AU, which sets the grid cell size in $r$ and $z$ to be about 0.025 AU across.


\subsection{Initial Models and Perturbations}

To study shock bores numerically, we generate spiral shocks by applying nonaxisymmetric perturbations to two equilibrium axisymmetric disks.
The initial disks are gravitationally stable and are only evolved for a few revolutions once the potential perturbation is applied.  The star/disk models are generated for the IUHG code using the self-consistent field (SCF) method \citep{hachisu86,pickett1996}.  Due to the nature of the SCF method, the EOS is restricted to a barotrope, so we assume our initial disks are polytropic, where
$ P= K \rho^\gamma $
and $K$ and $\gamma$ are constants; $\gamma$ being ratio of specific heats if the gas is interpreted as an isentropic ideal gas.  The central star is set to 1 M$_{\odot}$ and is removed from the grid, leaving only its potential.  The resulting inner holes for the two models are about 0.3 and 0.6 AU in radius.  The models are then evolved in an axisymmetric version of the three-dimensional code to allow for transient features introduced by removing the star to subside. 

For the simulations presented here, the initial models are similar to the high- and moderate-mass disks discussed in \citet{boley2005proc}. The high-mass disk is the same as the \citet{pickett2003} high-Q disk, but with its outer radius scaled to 6 AU.  This disk has a surface density profile of $r^{-0.5}$ and an initial temperature profile of $r^{-1.0}$, with a total mass of  0.14 M$_{\odot}$ inside 6 AU.  To relate this to a solar nebula model, suppose that $\Sigma(r)$ became $r^{-1.5}$ in the unmodeled disk outside 6 AU \citep[see][]{lissauer87} .  Then the total disk mass would be 0.82 M$_{\odot}$ out to 40 AU.  Although this may be unreasonably massive for a solar nebula, the disk is only meant to demonstrate the dynamics of shock bores.      The \citet{toomre64} 
$Q = c_s\kappa/\pi G\Sigma,$
where $c_s$ is the sound speed and $\kappa$ is the epicyclic frequency, is a parameter that indicates whether a disk is gravitationally stable.  If $Q < 1.5$, gravitational instabilities will set in \citep{durisengireview}.   With a minimum $Q=2$ near 5 AU, this disk is stable against gravitational instabilities.  

The moderate-mass disk contains 0.037 M$_{\odot}$ within 6 AU and also has an initial $\Sigma\propto r^{-0.5}$ out to 6 AU.  If the surface density profile breaks to $r^{-1.5}$ beyond 6 AU, the total mass contained within 40 AU is 0.21 M$_{\odot}$.  The initial minimum $Q$ is also approximately 2 at 5 AU.

Since shock bores can become very complex, especially when multiple jumps occur near each other, we stimulate a two-armed spiral wave.  This is done using two methods.  The first method forces spiral waves in the high-mass disk by adding a $\cos 2\varphi$ potential perturbation $\Phi_p$ centered near 5 AU with a radial FWHM of about 0.5 AU.  This potential perturbation has the form
\begin{equation}\Phi_p(r,~\varphi;~ t)= A \cos\left[2\left(\varphi - \Omega_p t\right)\right]\cos^2\left(\pi | r-r_p | \Delta R^{-1}\right), \label{potpert}\end{equation}
where $r_p = 5$ AU defines the pattern rotational speed $\Omega_p$, which is assumed to be Keplerian, $A$ is some scaling parameter, and $\Delta R=1$ AU.  
A benefit to this perturbation is that $\Phi_p$ sums to zero 
%
 %
and it creates a well-defined two armed spiral that reaches all the way down to the central hole (see Fig.\ \ref{fig3}) without changing the center of mass.   Since the perturbation is localized to 5 AU, the change in the potential far from 5 AU, which stimulates the waves, is due to the concentration of the mass at the potential minima.  In this way, the $\cos 2\varphi$ potential perturbation behaves like two stubby spiral arms that could be produced by gravitational instabilities \citep[see, e.g.,][]{bossdick}.  

For the moderate-mass disk, a different $\cos 2\phi$ perturbation is applied.  Because the outer portion of the modeled disk has low mass, the stubby spiral arms produced by equation (\ref{potpert}) do not have enough mass to generate strong shocks.  Instead, we place two 2.5 M$_{\mathrm{J}}$ corotating point masses at $r = 5.2$ AU in the midplane of the disk, separated azimuthally by $\pi$ radians.  Again, this method keeps the center of mass of the system at the center and creates two well-defined spiral arms, but it does allow for additional, small arms to form.    Although we are treating the perturbers as point masses, one should not interpret them strictly as protoplanets, because any such perturber would open a gap with in a few orbits \citep{bate2003}.  Instead, we envision these point masses to be transient clumps, formed by gravitational instabilities in a surface density enhanced ring. 

\subsection{Thermal Physics}

The behavior of shock bores depends on the EOS, as demonstrated in \S \ref{hstheory}. To test the predictions of the shock bore theory, the high-mass disk is evolved two ways: (1) with an energy equation that includes some cooling plus heating by bulk viscosity and the EOS for an ideal gas,  $P = (\gamma-1)\epsilon$, where $\epsilon$ is the internal energy density and $\gamma=5/3$, and (2) without an energy equation and an isothermal EOS, where $P(r,\varphi,z)=\rho(r,\varphi,z)T_{0}(r)$ and $T_{0}$ represents the initial axisymmetric midplane temperature.  The moderate-mass disk is evolved only with an energy equation.  When the energy equation is used, a gentle volumetric cooling rate, with a constant cooling time, is applied to the disk to partially balance shock heating.  Without any cooling, the disk heats up in less than a pattern period and surface waves are suppressed.  The cooling time for the high-mass disk is set to four pattern periods at the perturbation radius.  
A constant cooling time is also used for the moderate mass disk to balance shock heating, but with a cooling time of six pattern periods because the shocks are weaker.

\subsection{Fluid Element Tracer}

To investigate the parameters of the shock in both disks, we use a fluid element tracer, which is a fourth-order Runge-Kutta integration scheme that uses output files from the IUHG code to specify velocity fields and temperature, density, and pressure distributions.  For each time step in the integration, the grid information is linearly interpolated to the next output file.  The files are written out approximately every 1/400th of an orbit at the peak perturbation radius.  Since the data storage demands are extremely high to achieve this type of spatial and temporal resolution, we evolve the disks first at a low resolution, 128 aziumthal zones, until spiral shocks develop.  Once the shocks form a quasi-steady morphology in the midplane, which takes about one or two pattern rotations, we switch to high resolution, 512 azimuthal zones, and continue the simulations for a little more than half a pattern period.  The isothermal calculation is only done at low resolution, because the low resolution is sufficient to demonstrate the general behavior of shock bores under isothermal conditions. The model parameters are summarized in Table \ref{models}.

\section{Results}\label{results}

In the simulations with heating and cooling ("Energy eq." in Table 1), the disk perturbations result in shock bores along the spiral arms.  The jumping material creates very large waves that extend to nearly twice the height of the disk, and the only steady features in the simulations are the presence of the shock fronts and the shock bores.  The shock bores vary in strength, and the nonlinearity of the waves creates transient features. 

 These bores and breaking waves created by spiral shocks in disks are fully three-dimensional; looking at only a single projection of a wave's morphology and its corresponding gas flow can be  both confusing and misleading.  In the discussion below, we present two two-dimensional cuts, $r$-$z$ and $\varphi$-$z$ at different disk positions.   When considered together, they form a comprehensible picture of waves in the disk.   In the projections, we use contours to delineate density and shock heating by artificial viscosity,  and we use arrows to represent gas velocity vectors.   It is important to remember that one cannot trace fluid element trajectories by connecting gas velocity vectors in any one projection. 

\subsection{High-Mass Disk}
\begin{figure}
\plotone{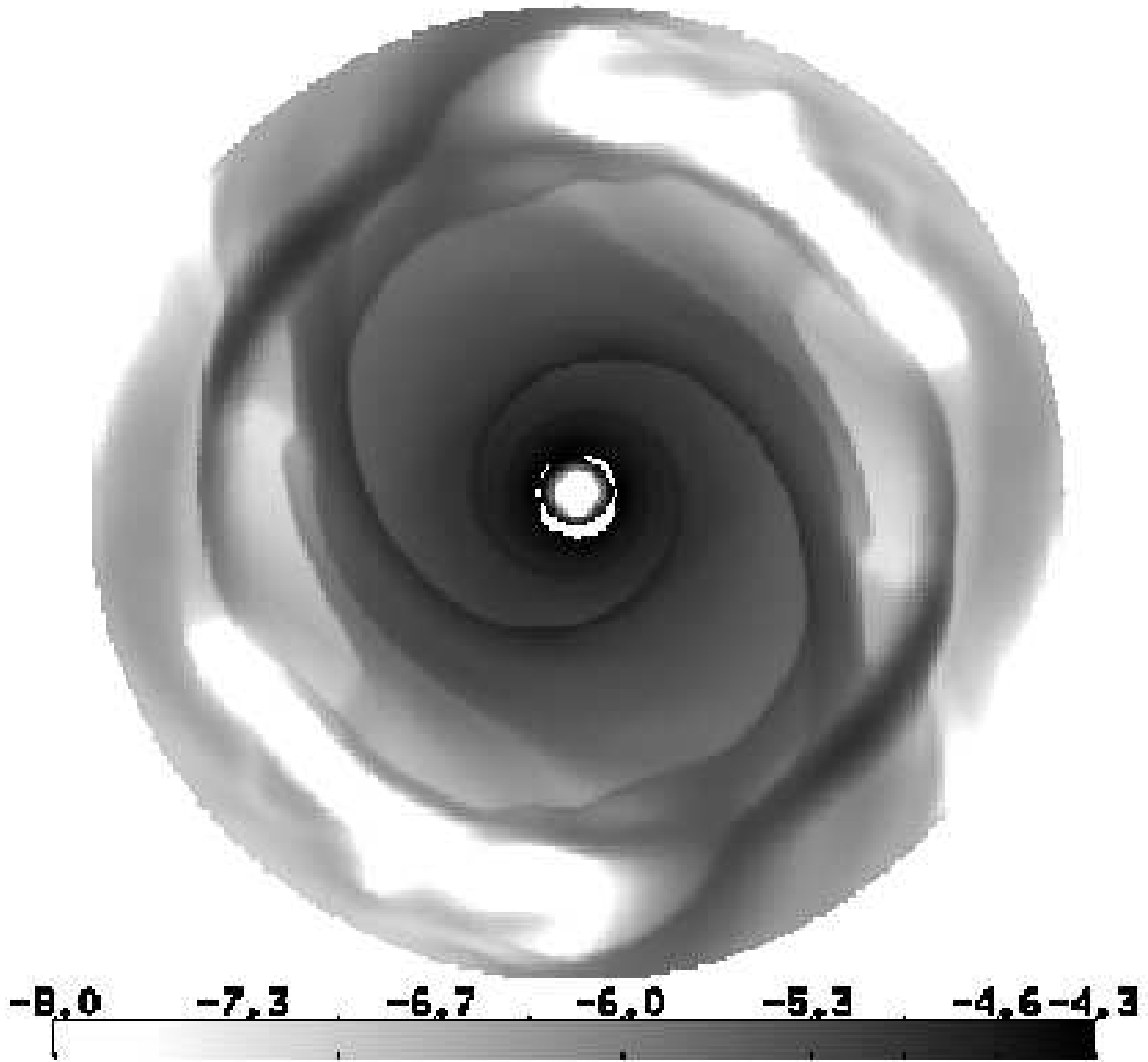}
\caption{Midplane logarithmic density gray scale for the high-mass disk.  The units for the gray scale are given in code units.  The conversion to real units is 3.54(-4) g cc$^{-1}$ per code unit.  Flow in this diagram is counterclockwise.  The total region shown has a radius of 6.4 AU.}\label{fig3}
\end{figure}

\begin{figure}
\plotone{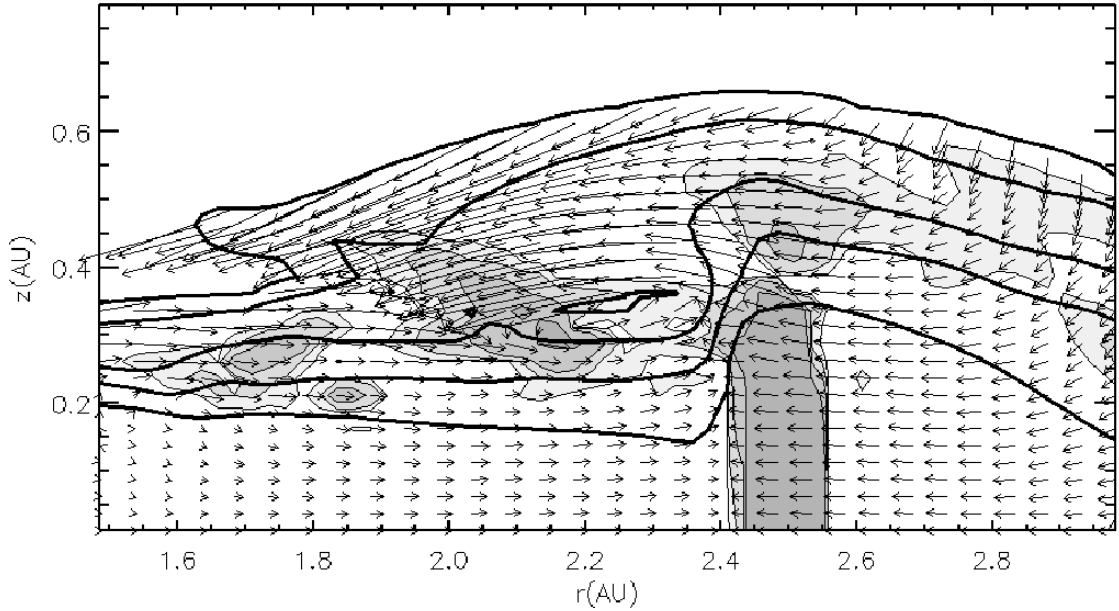}
\caption{A radial cross section portraying the shock at $r=2.5$ AU.  The thick lines are density contours corresponding to 3.5(-12), 3.5(-11), 3.5(-10), 7.1(-10), and 1.4(-9) g cc$^{-1}$.  The grey, shaded regions indicate shock heating corresponding to 5.6(-9), 2.2(-8 ), 9.0(-8), and 3.6(-7) ergs cc$^{-1}$ s$^{-1}$.  The arrows show the velocity of the gas with each component scaled to its appropriate axis.}\label{fig4}
\end{figure}

The cross sections showing the spiral shock at $r=2.5$ AU for the simulation with heating and cooling presents the cleanest morphology, so we will examine it first.  Figure \ref{fig3} is the midplane density gray scale of the disk at the time that the following cross sections are shown and is used as a point of reference.  Figure \ref{fig4} shows the velocity vectors of the gas in an $r$-$z$ cut, which corresponds to 8 o'clock in Figure \ref{fig3}.  The arrows represent the velocity components of the gas, with each component scaled to its axis; the thick contours represent the density structure, and the light contours with grey fill represent shock heating.  Near the midplane, where fluid elements are weakly affected by the shock bores, fluid element trajectories are roughly arranged in streamlines, creating the oval distortion seen in Figure \ref{fig3}, with the apastron near the spiral shock.  This is indicated by the gas velocity vectors near 2.4 AU in Figure \ref{fig4}.  For $r>2.4$ AU, the gas at all altitudes has already passed through a spiral shock at a larger radius and is flowing radially inward.  The downward moving, high-altitude material that seems to appear with no origin is passing through this cross section and is part of a coherent flow.  For $r<2.4$ AU, the low-altitude material ($z<0.3$ AU) represents the outwardly moving pre-shock flow, while the mid- to high-altitude disk material has already passed through a spiral shock and has now developed into a large breaking wave.   The breaking wave produces high-altitude shocks over the pre-shock flow and a large vortical flow, which appears to have some effect on the pre-shock flow even at low-disk altitudes.  The disk height reaches a maximum near the spiral shock and is slightly less than twice its unperturbed height. We discuss the height of the jump more quantitatively in \S \ref{discussion}.
 
\begin{figure}
\plotone{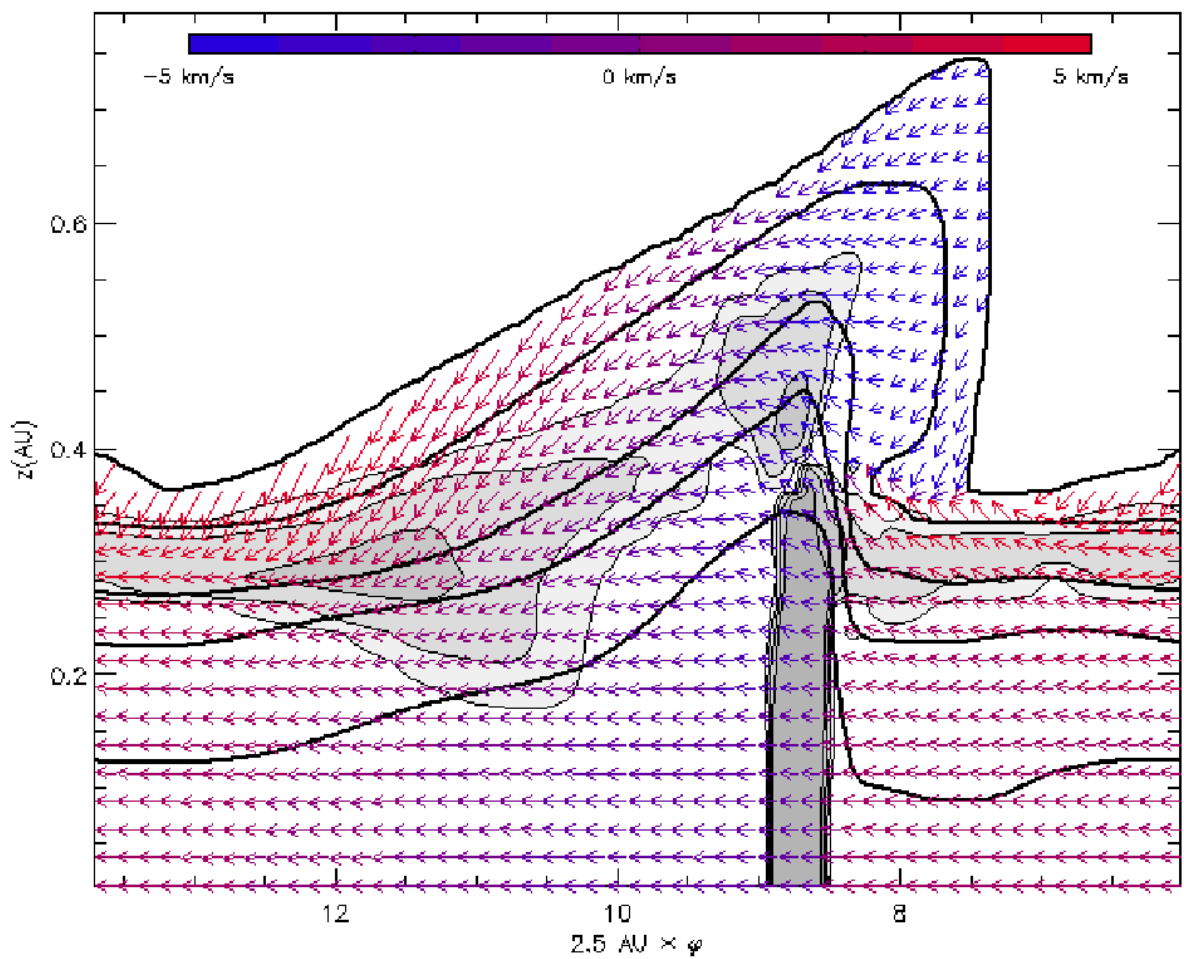}
\caption{Cylindrical cross section at $r=2.5$ AU for the same time shown in Figs.\ \ref{fig3} and \ref {fig4}.  The density and heating contours are the same as in Fig.\ \ref{fig4}. The arrows represent the flow of the gas in the frame of the potential perturbation and the color of the arrows represent the radial flow of material through this cross section (blue toward the reader or central star and red away). This cross section is shown from the perspective of an observer at the central star, and $\varphi$ is measured counterclockwise from the 3 o'clock position in Fig.\ \ref{fig3}.}\label{fig5}
\end{figure}

 Figure \ref{fig5} is a $\varphi$-$z$  cross section at $r=2.5$ AU, where $\varphi$ is measured counterclockwise from the 3 o'clock position in Figure \ref{fig3}, with velocities shown in the reference frame of the spiral wave, and complements Figure \ref{fig4}.  In this figure, the arrows are colored (in the online version) to represent the radial flow of the gas as seen by an observer positioned at the star.  Before the shock, the gas is moving radially outward, and after the shock, it is moving radially inward.  In addition, the mid- to high-altitude gas starts its vertical expansion at that interface.   The sharp wall of gas at $2.5\mathrm{\ AU}\times\varphi=6$ AU is formed by gas  that jumped at a larger radius and is now falling inward through this cross-section.  As this material crashes back onto the disk, it creates high-altitude shocks that result in the expansion of the high-altiude gas as the post-shock region transitions into the pre-shock region of the next arm; the gas begins to move radially outward again due to the roughly elliptical orbits of the gas.  At a later time, not shown, the wall is still present, but it is less distinct, and the gas forms a tube similar to a breaking surface wave.  The structure of shock bores is time dependent and transient.\footnote{An animation showing the time-dependence of shock bores is made available at http://westworld.astro.indiana.edu/. Download the file "Shock wave" under the Movies link.}

\begin{figure}
\plotone{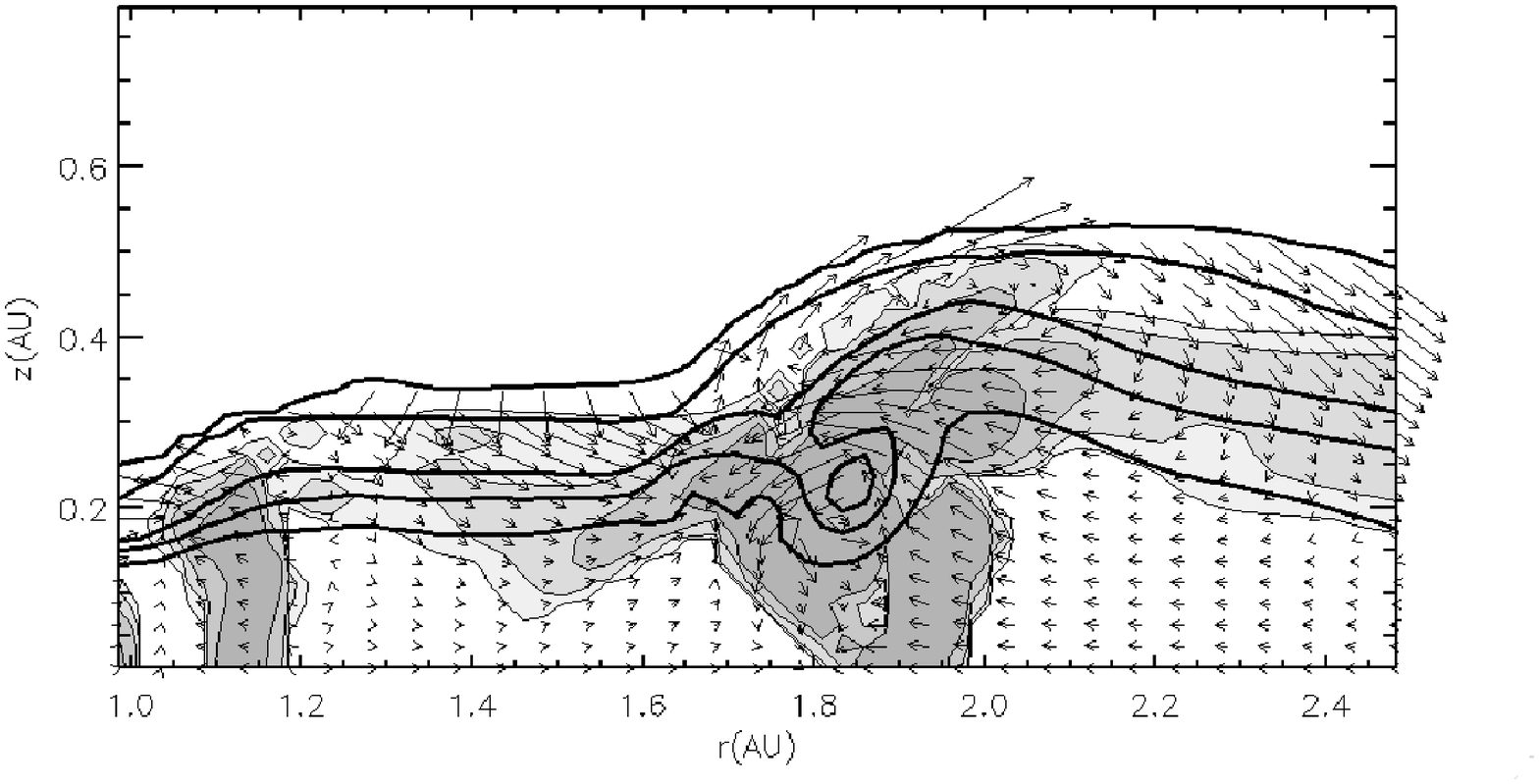}
\caption{Same as Fig.\ \ref{fig4}, except cutting across the shock at $r=1.9$ AU.}\label{fig6}
\end{figure}

\begin{figure}
\plotone{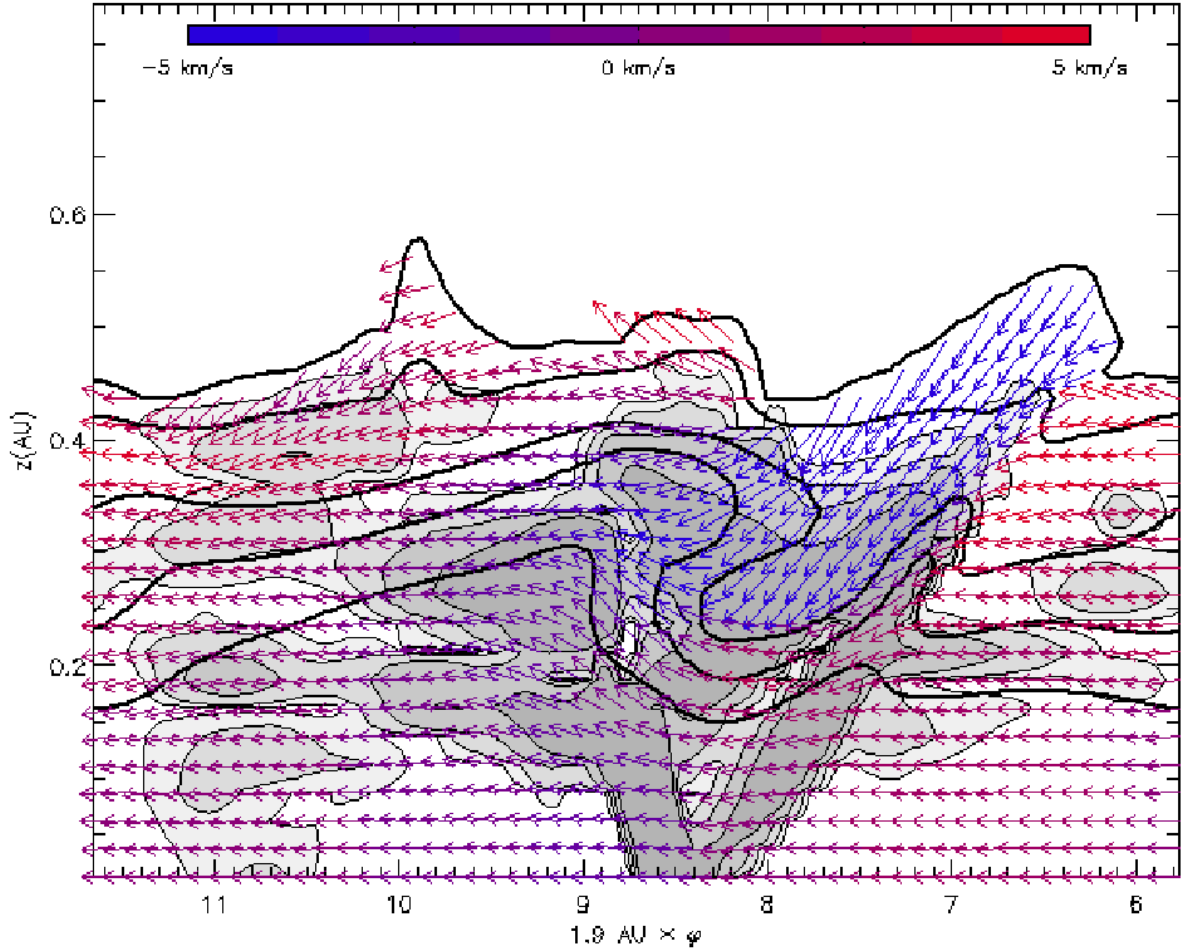}
\caption{Same as Fig.\ \ref{fig5}, except for $r=1.9$ AU.  The shock morphology is much more complex and the shock bore cannot easily be defined by a single shock.  Note the material, which jumped at a larger radius, plunging down into the disk between 7 and 9 AU.}\label{fig7}
\end{figure}

Figure \ref{fig6} shows the region 0.6 AU inward at about 6:30 o'clock in Figure \ref{fig3}.  The breaking wave drastically affects the high density gas flow.  A clear shock front cannot be defined for our coarse three-dimensional grid, but strong shocks are everywhere in the wave.  A large vortical flow reaches down to the midplane at $r=1.8$ AU; another vortical flow is present in the high-altitude gas at $r=2.0$ AU.  These circulation cells are clearly related to the shock bores and waves.  In the corresponding cylindrical cross section, Figure \ref{fig7}, the complexity of the shock structure is again highlighted, and it is difficult to distinguish the shock bore, which is at about $1.9\times\varphi\rm\ AU=9\ AU$, due to the strong disturbance by gas that jumped at a larger radius and is plunging into the disk between  $1.9\times\varphi\rm\ AU=7$ and 8 AU.

 On the other hand, the shock bore  shown in  Figure \ref{fig8} for $r=3.1$ AU at about 9:30 o'clock in Figure \ref{fig3}, which is near the Lindblad resonance of the imposed $m=2$ perturbation, shows the development of a very large wave that has yet to break onto the disk's surface.  The cylindrical cross section for this location, Figure \ref{fig9}, indicates that the wave is associated with a very sharp wall, similar to the 2.5 AU morphology.  
 
These figures show that, at all disk radii portrayed, the disk material at high altitudes is moving differently from the material near the midplane, shocks are present at all disk altitudes, and the shock bores lead to large breaking waves, which create extensive vortical flows in the disk.  The effect that these bores and waves have on the disk surface is emphasized by the isodensity surface contour shown in Figure \ref{fig10}.

The high-mass disk was also evolved using an isothermal equation of state using the same potential perturbation.  Figure \ref{fig11} shows a $\varphi$-$z$ cross section for a shock near $r=2.5$ AU.   Instead of expanding at the shock, the gas compresses slightly and the height of the disk remains about the same.   We speculate that the small peak of downward-moving, low density material just before the spiral shock is a remnant of transient features created by suddenly switching the EOS of the disk and by forcing a strong disk perturbation, and it is not part of the shock bore. The vertical compression of the gas at the shock front is in agreement with equation (\ref{accelpart2}) because $J_f \le 1$ for an isothermal shock as given by equation (\ref{jfiso}).  In an isothermal gas disk, a spiral shock will behave like a spiral density wave if self-gravity is unimportant, or it will compress vertically along the shock front if self-gravity is important.  

\subsection{Moderate-Mass Disk}\label{resultsmoderate}

The moderate-mass disk portrays a different shock morphology from the high-mass case.  The shocks are weak, and, therefore, the shock bores are weaker and do not occur globally.  The wave structure in the disk may be more closely related to $f$-modes, as described by Lubow \& Ogilvie (1998).  Figure \ref{fig12} shows the $r$-$z$ cross section for the shock at $r=2.7$ AU, which is at the same relative position as the 2.5 AU cross section of the massive disk.  There is surface corrugation and a weak shock bore at the spiral shock, but a breaking wave does not form. Figure \ref{fig13}, the cylindrical cross section, shows that a strong vertical wall formed by inward falling material is, for the most part, absent. Nevertheless, the gas does expand vertically after entering the shock.  The moderate-mass disk also exhibits very complex wave interactions that we are still investigating and hope to discuss in a future paper. 

\section{DISCUSSION}\label{discussion}

\subsection{Shock Bores}

Each of the three simulations shows different shock and wave morphologies.   In the high-mass disk, strong shocks are produced, which lead to the formation of shock bores over most of the spiral wave.  The spiral waves in the isothermal calculation propagate principally as spiral density waves.  The moderate-mass disk only shows a bore-like morphology near $r = 2.7$ AU, and the disk has weak spiral shocks.   It appears that shock bores are the extreme nonlinear outcome when $f$-modes develop into strong spiral shocks.  To understand how well the shock bore theory describes this nonlinear regime, we use the fluid element tracer to quantify shock parameters.

\begin{figure}
\plotone{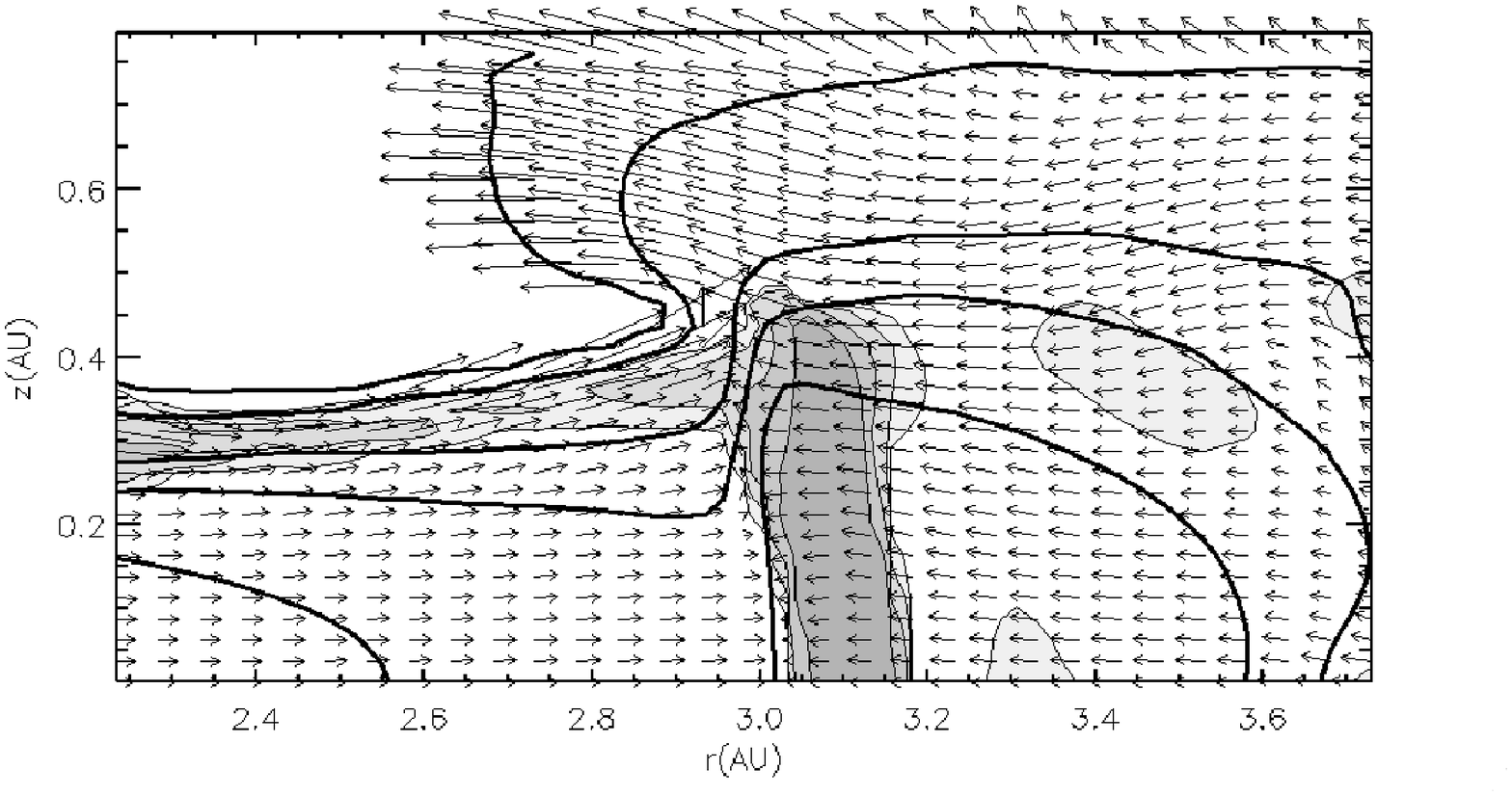}
\caption{Same as Fig.\ \ref{fig4}, except cutting across the shock at $r=3.1$ AU.}\label{fig8}
\end{figure}

\begin{figure}
\plotone{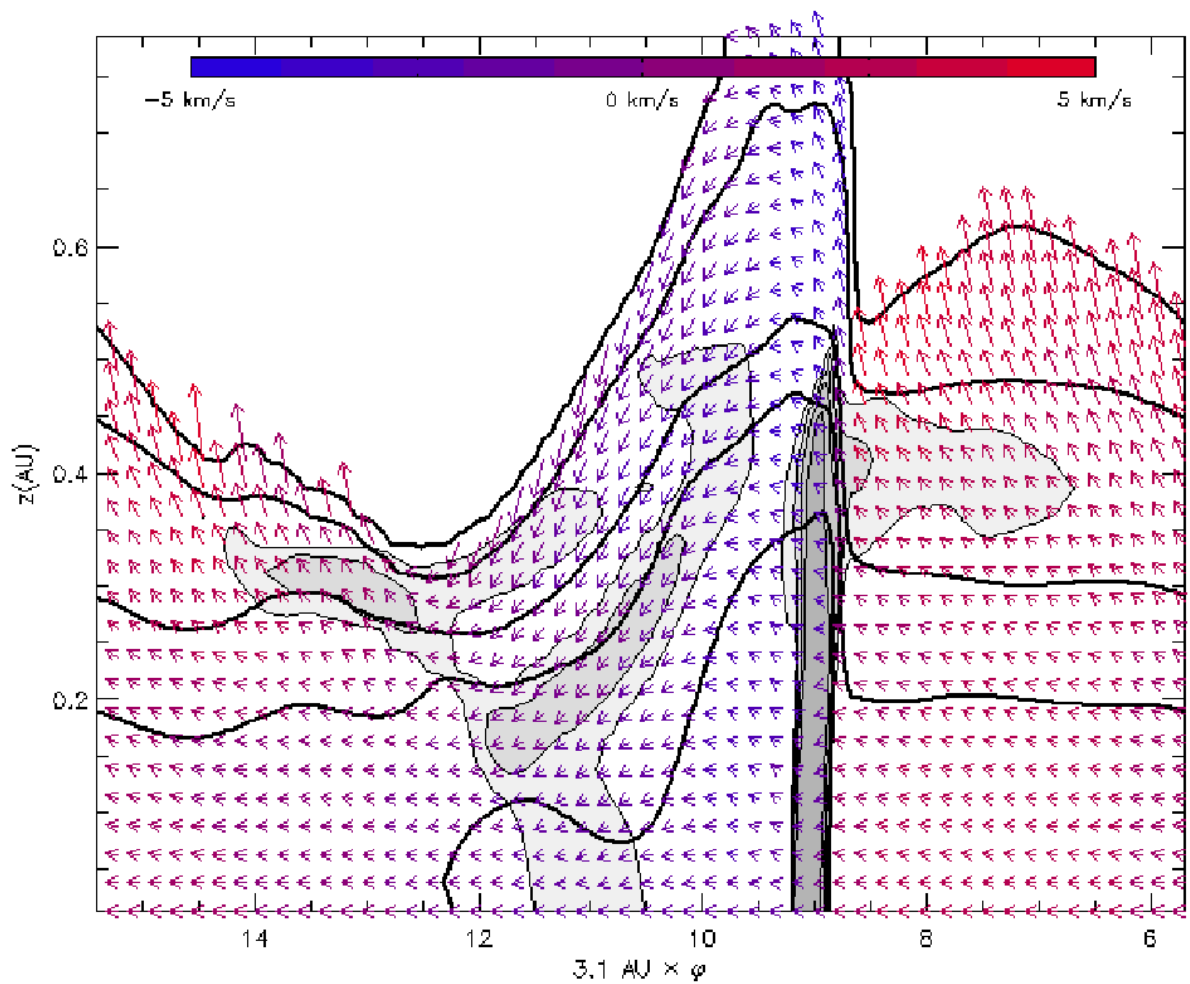}
\caption{Same as Fig.\ (\ref{fig5}), except for $r=3.1$ AU.  Even though this cross-section is the closest to the corotation radius, a strong shock and shock bore are still observed.  This is probably due to the large radial motions of the fluid elements.}\label{fig9}
\end{figure}

\begin{figure}
\plotone{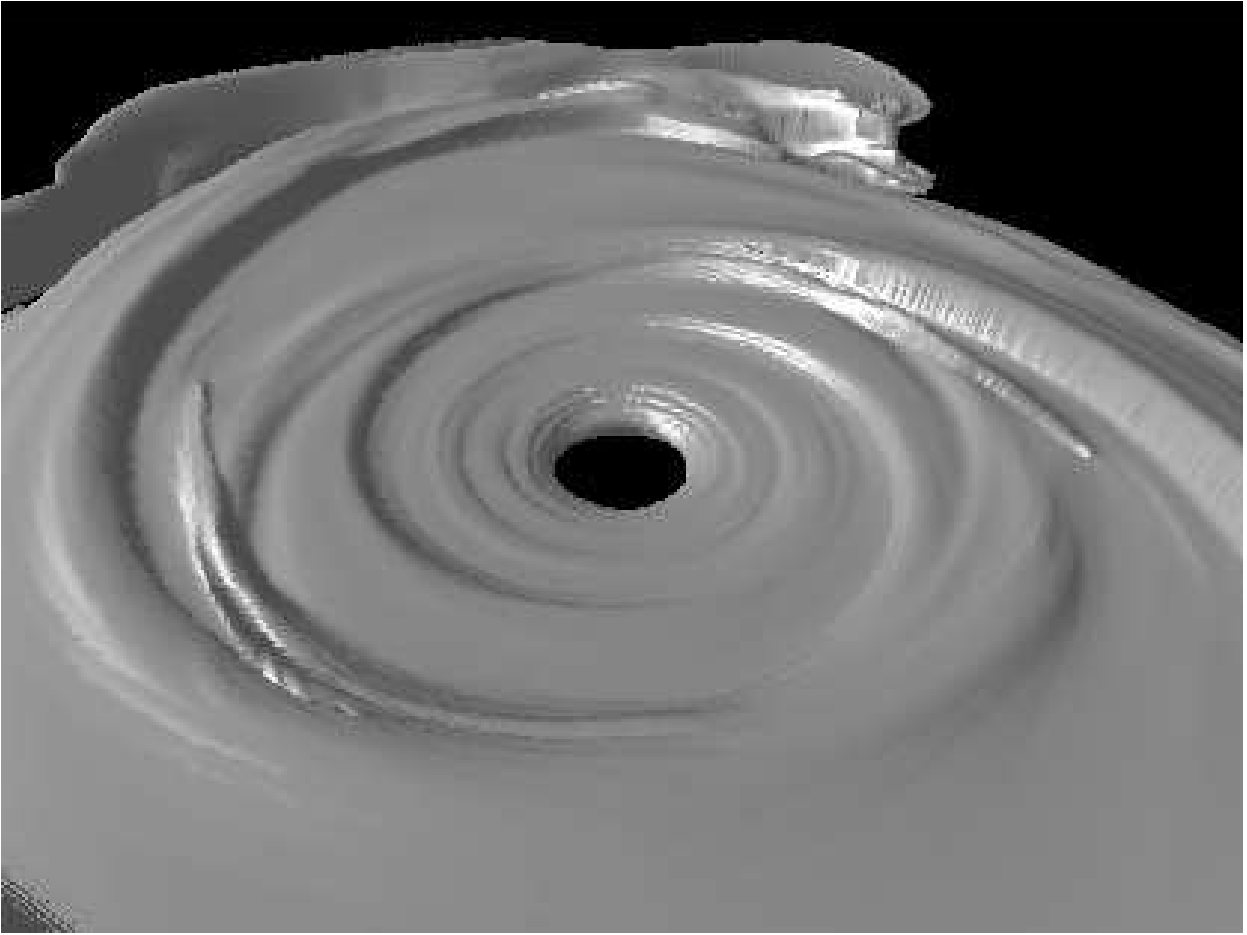}
\caption{Isodensity surface contour for $\rho=3.54(-10)$ g cc$^{-1}$ shown at the same time as Figs.\ \ref{fig4}-\ref{fig9}. The view is from above the disk looking from about 5 o'clock to 11 o'clock.  The surface contour routine is unable to show breaking waves clearly.}\label{fig10}
\end{figure}

\begin{figure}
\plotone{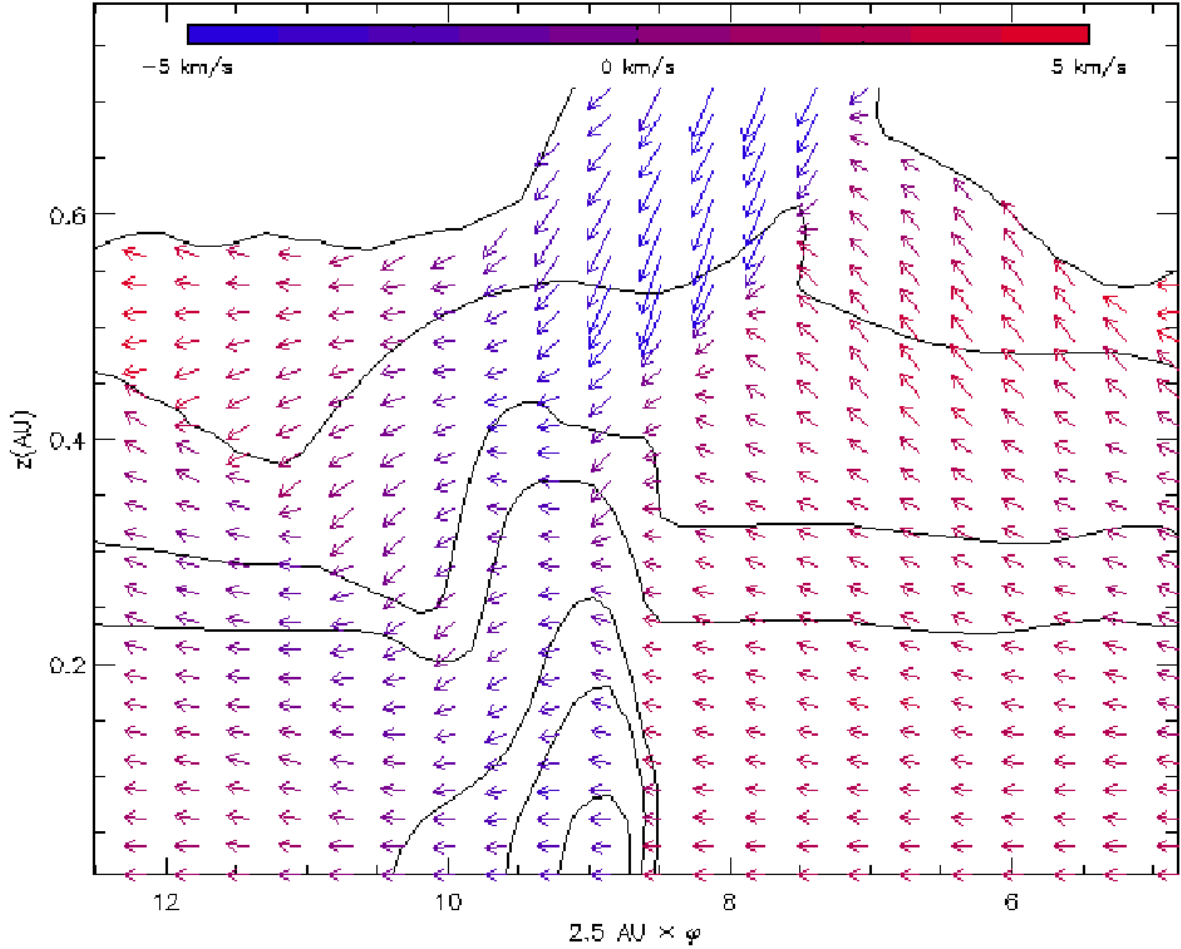}
\caption{Cylindrical cross section at $r=2.5$ AU for the isothermal high-mass disk simulation.  The density contours are the same as in Fig.\ \ref{fig5}.  The shock front does not cause rapid expansion in the post-shock region, but instead causes a compression.  We speculate that the small peak of downward-moving, low-density material just before the spiral shock is a remnant of transient features created by 
suddenly switching the EOS of the disk and by forcing a strong disk perturbation. Although this wave is not associated with the compression caused by the shock bore, it is probably responsible for the slight undulatory morphology in the post-shock region.}\label{fig11}
\end{figure}

\begin{figure}
\plotone{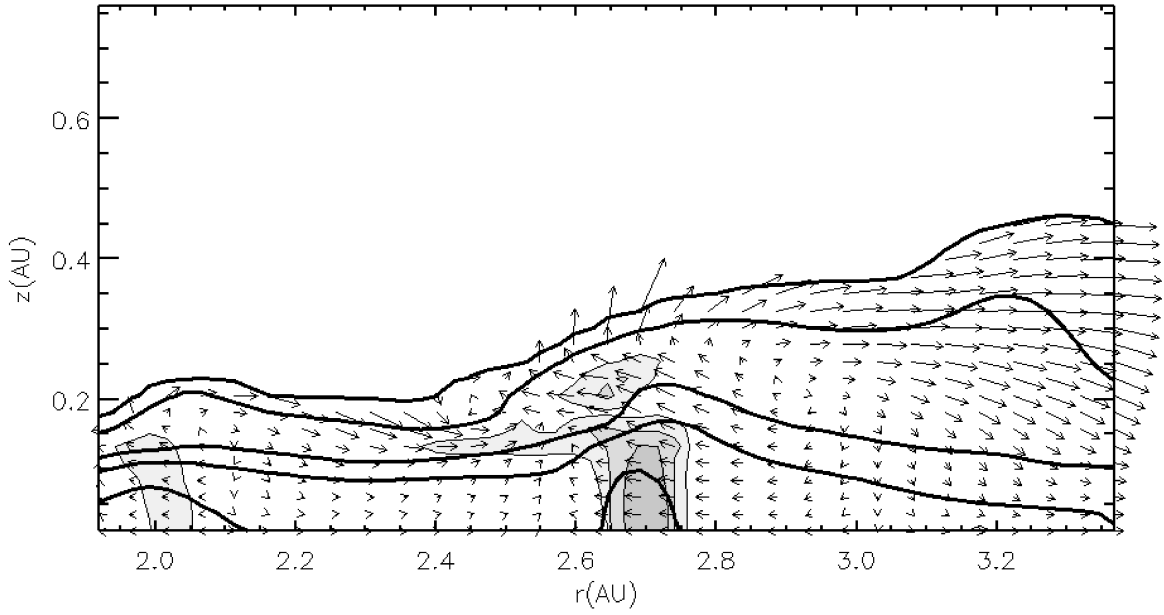}
\caption{Same as Fig.\ \ref{fig4}, except for the moderate-mass disk cutting across the shock at $r=2.7$ AU.}\label{fig12}
\end{figure}

\begin{figure}
\plotone{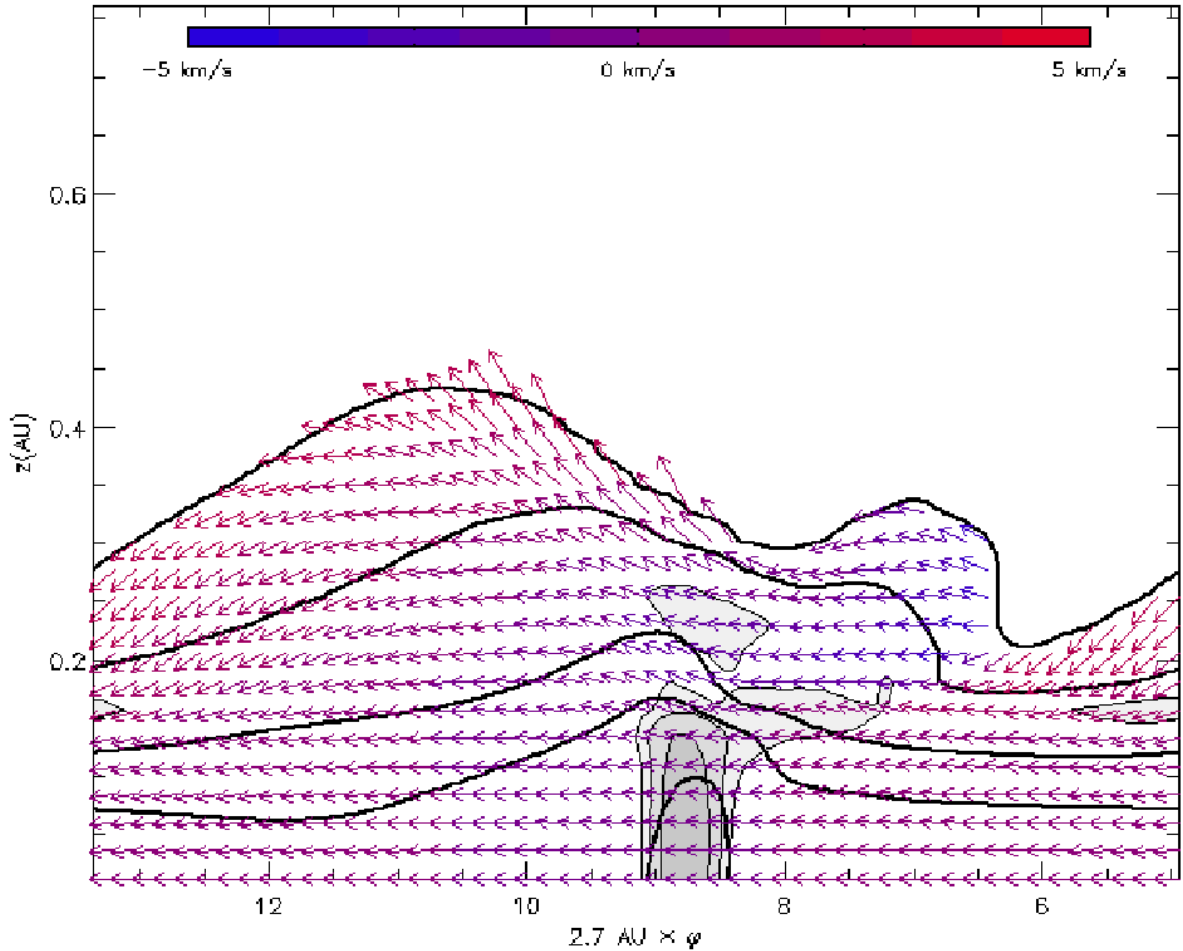}
\caption{Same as Fig.\ \ref{fig5}, except for the moderate-mass disk at $r=2.7$ AU.  The shock is weaker but still strong enough to induce a shock bore.  In addition, the peak at 11 AU, which is moving radially outward, is not obviously associated with the shock bore.  The nonlinear dynamics of waves in these disks is complex; the shock bores are not necessarily the whole story. }\label{fig13}
\end{figure}

\begin{figure}
\plotone{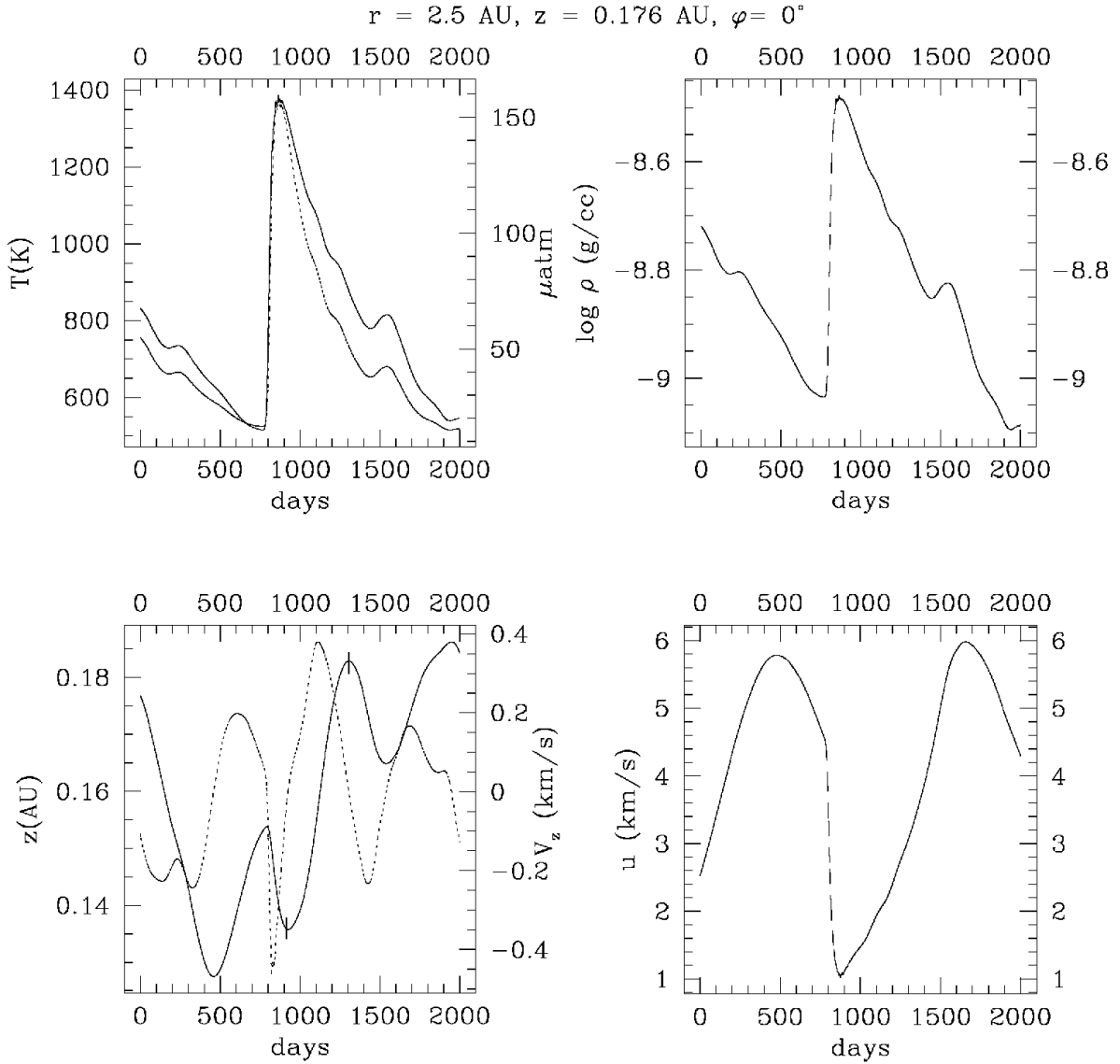}
\caption{Example of a thermal history for a fluid element starting at $r=2.5$ AU, $z = 0.176$ AU, and $\varphi = 0^{\circ}$.  {\it Top left:} Temperature (solid line) and pressure (dotted line) histories.  {\it Top right:} Density history. {\it Bottom left:} Disk altitude trajectory (solid line) and $V_z$ (dashed line).  Although the vertical velocity does become negative and the vertical altitude reaches a local minimum while it is going through the shock, it quickly expands with a very sharp change in the vertical velocity as soon as it is in the post-shock region.  Roughly, the shock bore is between the hash marks. {\it Bottom right:} Shock normal velocity, with respect to the global spiral shock, in the frame of the shock.  }\label{fig14}
\end{figure}

\begin{figure}
\plotone{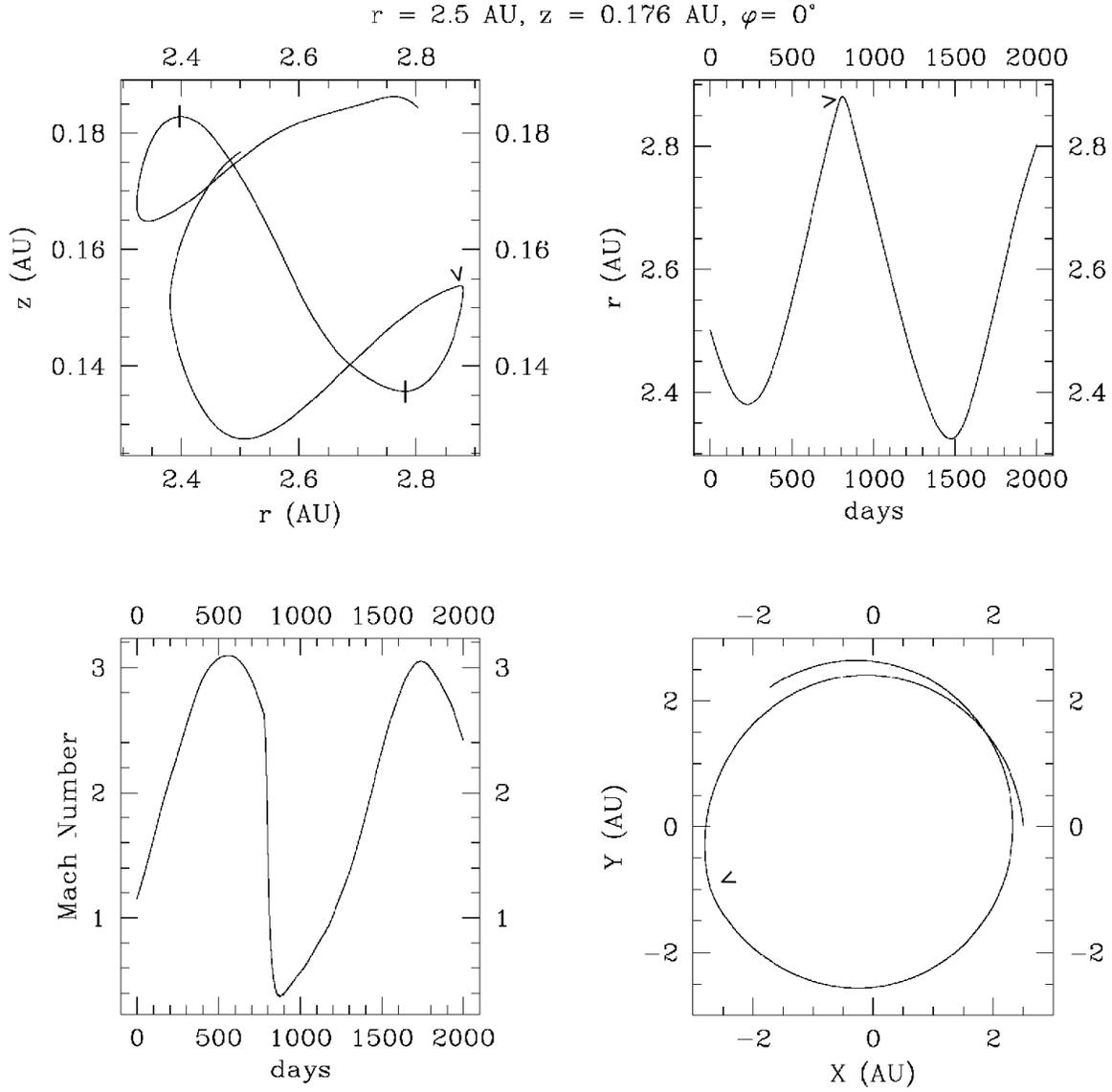}
\caption{Complement to Fig.\ (\ref{fig14}), showing the fluid element trajectories. {\it Clockwise:} $r$ vs.\ $z$, $r$ vs.\ time, Mach number vs.\ time based on the $P$, $\rho$ and $u$ information presented in Fig.\ (\ref{fig14}), and the projection of the motion onto the midplane in Cartesian coordinates. The open arrow heads show where the shock begins.  Roughly, the shock bore is between the hash marks.}\label{fig15}
\end{figure}

Figure \ref{fig14} shows a sample thermal history for a fluid element starting at $r = 2.5$ AU, $z$ = 0.176 AU, and $\varphi$ = 0$^{\circ}$ in the high-mass disk.  The fluid element is placed in the disk when the shocks first become strong, at the same time that the grid is switched to high resolution, and the fluid element is followed for half a pattern period.  The shock is seen clearly in the density and temperature/pressure plots.  It is also noticeable in the $u$-profile, where $u$ is the shock normal speed in the frame of the pattern. To estimate $u$, it is assumed that the pitch angle $i$ of the spiral arms is constant, nearly 20$^{\circ}$ for these calculations.  With this assumption,
\begin{equation}u = v_{r}\cos i + (\Omega-\Omega_p)r \sin i\label{pitch}.\end{equation}
Although this will not precisely measure $u$, and therefore the Mach number, it will, for most of the shocks in the disk, give us a reasonable value. 
Figure \ref{fig15} indicates the Mach number as well as the trajectory of the fluid element corresponding to Figure \ref{fig14}.  Even though the fluid element starts at $r=2.5$ AU, it is transported outward to 3.1 AU (Fig.\ \ref{fig8}) before it encounters the shock.  The Mach number for the shock in Figure \ref{fig8} is around 2.6, which roughly agrees with the change in density from equations (\ref{jf2}) and (\ref{scaleheights}). The resulting change in the disk height should therefore be about 1.7, in the non-self-gravitating limit, or 1.6 assuming a $q =10$.  The actual vertical jump of a factor 1.5 is similar.    Similar calculations indicate that the Mach number for the spiral shock shown in Figure \ref{fig4}, the radius where the fluid element is originally launched, is about 2.7.  The jump in disk height again appears to be about 1.5-1.6.

Even though the jump morphology in the $r=1.9$ AU cylindrical cross section of the high-mass disk is unclear, the shock bore description of spiral shocks predicts the new height for the inner disk fairly accurately.  The Mach number for the shocks at $r$ = 1.9 AU is also about 2.7.  According to equation (\ref{scaleheights}), the new disk height should also be about 1.6-1.7 times higher for $q\rightarrow\infty$ than the original height. The initial height of the disk is about 0.3 AU at $r$ = 1.9 AU and the height of the disk is about 0.5 AU in Figure \ref{fig7}, which matches the prediction well. 

The simple shock bore picture may be complicated by other wave dynamics near the Lindblad resonance at $r=3.1$ AU (see Figs.\ \ref{fig8} and \ref{fig9}), but shock bore theory still seems to predict the behavior of the shocks well in this region.  Along the spiral shock, a bore is clearly seen.  The fluid element tracer indicates that the Mach number for the shock at mid-disk altitudes is near 2.6, as stated above.  The Mach numbers probably remain high at these radii because the fluid elements are on more highly elliptical orbits than in the inner radii, and the radial component to $u$ becomes important, see equation (\ref{pitch}), for these small pitch-angle spirals.  

For the moderate-mass disk, the fluid element tracer indicates that the spiral wave forms a distinct shock near $r=2.7$ AU.  Since the shock is weak, $u$ is unreliable, so we put the measured density change into equation (\ref {rhofull}) to find a Mach number of about 1.7.  Assuming self-gravity is negligible, we expect that the maximum height of the jumping material to be about 1.3 times the initial scale height, which is fairly consistent with Figure \ref{fig12}. 

\begin{figure}
\plotone{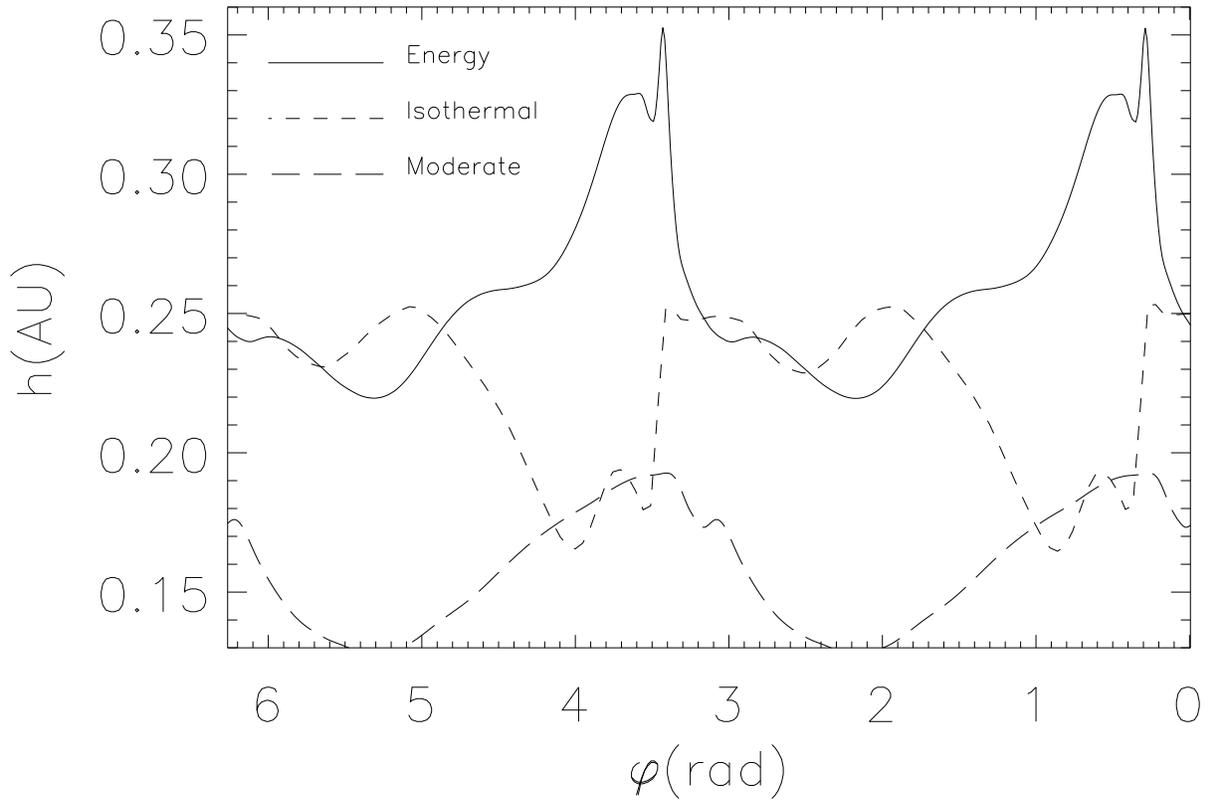}
\caption{Scale height of each disk vs.\ azimuth.  The high-mass disk, both energy and isothermal, are plotted for $r=2.5$ AU and the moderate-mass disk is plotted for $r=2.7$ AU.  The shocks all occur between about 2 and 4 rad.}\label{fig16}
\end{figure}

In discussing the changes of height in the disk, we have so far been limited to eyeballing the density contours, which can be a misleading indication of the size of the bore. Let us define a local disk scale height by%
\begin{equation}h=\int_0^{\infty}\mathrm{d}z~\rho(r,\varphi,z)/\rho_{\mathrm{mid}}(r,\varphi).\end{equation}
Figure \ref{fig16} plots this scale height for all three simulations at a similar radius.  The scale heights in the pre-shock region for the energy and isothermal calculations are similar, but the isothermal disk shows a decrease in the scale height after the shock, while the energy equation disk shows an increase.  The change in disk scale height for the energy equation is about 1.5, compared to 1.6 or 1.7 predicted by equation (\ref{scaleheights}) for the shock at this radius.  The moderate-mass disk has a more gradual change in scale height than the other calculations, because of its weak shocks.  The scale height changes by a factor of 1.3, close to what is predicted by equation (\ref{scaleheights}).   

\subsection{Processing and Mixing}

Large scale turbulent flow is present in both disks, which could result in the mixing of disk material radially and vertically.   In addition, the high-altitude disk material can be moving at a much greater speed, and even in a different direction, than the mid- and low-altitude material.  This makes it is possible to have a {\it slip surface} or {\it vortex sheet} at a Mach intersection, i.e., an intersection of two parallel flowing streams with different Mach numbers \citep[e.g.,][]{massey1970}.  This will provide additional turbulence in the disk on a small scale and could influence the evolution of solids in the disk, e.g., size-sorting of chondrules \citep[e.g.,][]{cuzzi2001}.  Unfortunately, we cannot model this with our simulations since the typical cell size is $\sim 10^{11}$ cm, which is too large to study turbulence on the Kolmogorov scale at which chondrules would be size-sorted \citep{cuzzi2004}.  Nevertheless, the energy contained in any shock bore, or even a strong wave, is enough to provide the necessary chondrule size-sorting turbulence in a disk \citep[see][]{boley2005proc}.  

\begin{figure}
\plotone{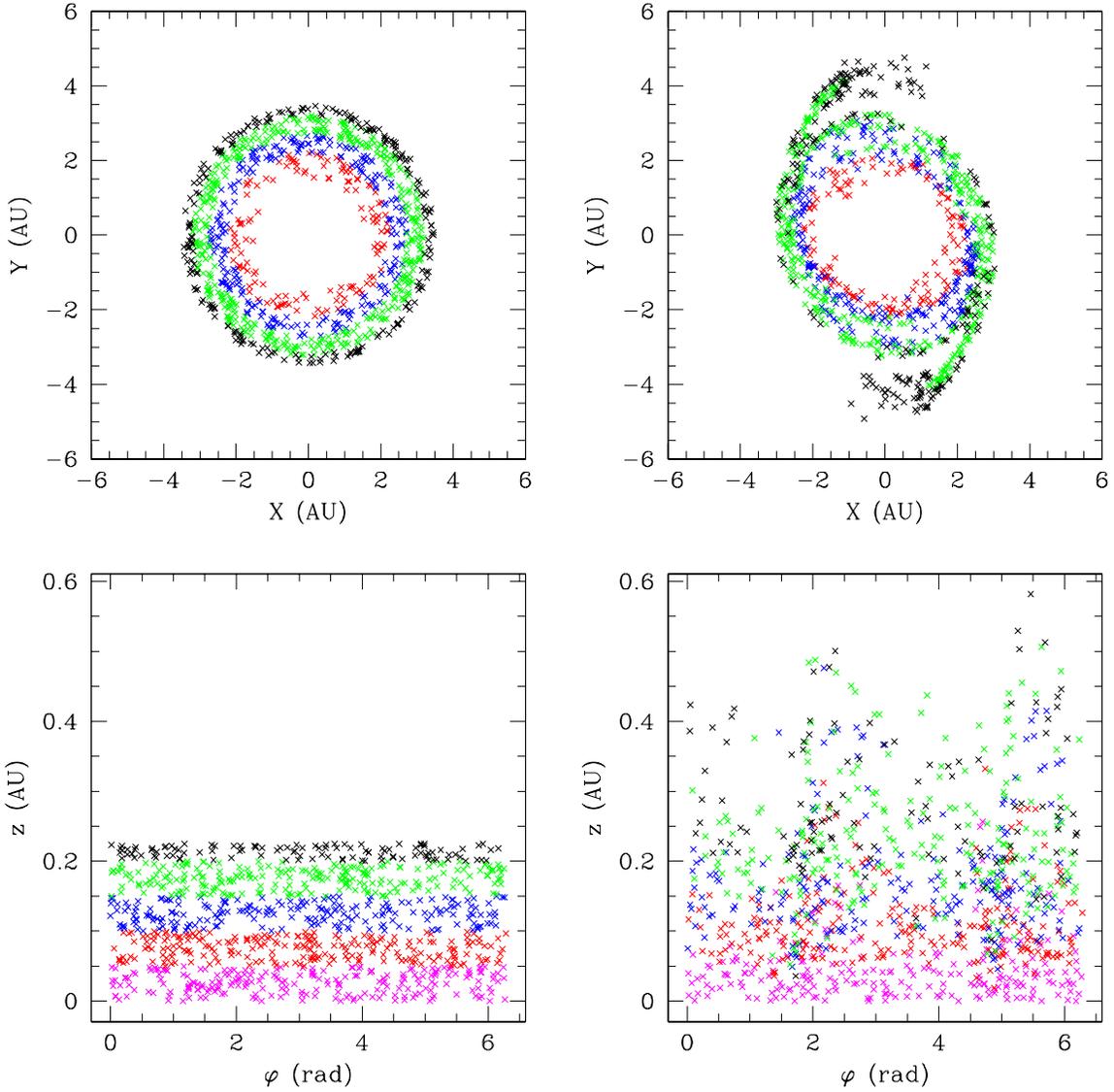}
\caption{Results of tracing 1000 randomly selected fluid elements within an annulus for half a pattern period.  {\it Top left:} Initial radial and azimuthal fluid element positions.  {\it Top right:} Position of the same elements after half a pattern period. {\it Bottom left:} Initial vertical distribution. {\it Bottom right:} Final vertical distribution.  Note that even material near the midplane is starting to show signs of stirring.}\label{fig17}
\end{figure}

We can look at the large-scale vertical and radial mixing/stirring produced by the spiral shocks and shock bores by using the fluid element tracer.  Figure \ref{fig17}  shows the results of tracking 1000 fluid elements for half a pattern period.  The fluid elements are randomly distributed initially between $r = 1.44$ and 3.36 AU and between $z =0$ and 0.24 AU.  At this stage in our analysis, we are unable to distinguish between {\it mixing} and {\it stirring} of nebular material.  Figure \ref{fig17} demonstrates, however, that shock bores do at least produce significant stirring over tenths of an AU in only about 5.5 yr.  Presumably, if we had a longer stretch of data, the stirring would involve more of the midplane material and be more extensive radially.  We are currently running simulations for a longer time period, and we hope to address mixing soon.  Moreover, we predict that the corotation radius will act like a barrier for the mixing, where material inside corotation can be mixed/stirred, material outside corotation can be mixied/sitrred, but the mixing/stirring will not cross the corotation radius because bores will be absent there.

Episodic mixing and stirring of the nebula could be driven by spiral shocks that form from repetitious clump or arm formation over several million years.   This would also provide short-lived shocks that could process solids throughout the nebula \citep[e.g.,][]{bossdick,boleyppv}, and may generate disk turbulence \citep{boley2005proc}.  Waves and shock bores could represent the major mechanism by which solids are processed and packaged into their parent bodies in protoplanetary disks.

\subsection{Shock Bores and Convection}

Convection has the potential to significantly alter the evolution of protoplanetary disks in several ways.  It could be a source for angular momentum transport \citep{lin_papaloizou1995}, a source for small-scale turbulence that leads to size-sorting of solids \citep[e.g.,][]{cuzzi2001}, a mechanism for rapidly cooling disks \citep{boss2002}, or some combination of these and other effects.   \citet{ruden_pollack1991} showed, using simple approximations, that convection should occur in disks when the vertical optical depth $\tau \gtrsim 1.78$ for a $\gamma=7/5$ and an opacity law $\kappa\approx\kappa_{0}T^2$.  Using the same approximations, if $\gamma=5/3$, convection should occur when $\tau\gtrsim5.3$.  However, these values for $\tau$ are only valid when ignoring all other dynamical effects.  Thermal convection is normally defined in terms of buoyancy due to thermal gradients operating in a {\it hydrostatic} background. Futhermore, as discussed by \cite{lin_papaloizou1995}, convection should be suppressed in disks when the effective $\alpha > 10^{-2}$ for an $\alpha$-disk accretion model \citep{alphadisk}.   \citet{mejiaphd} found effective $\alpha > 10^{-2} $ in her simulations undergoing gravitoturbulence.

To search for possible convective flows, we look for negative entropy gradients, i.e., where 
$\partial (P/\rho^{\gamma})/\partial z < 0$, instead of superadiabatic gradients, where $\nabla = \partial \ln T/\partial \ln P > \nabla_a = 0.4$. Here, $\nabla$ is the actual temperature gradient and $\nabla_a$ is the adiabatic gradient (0.4 for $\gamma=5/3$).  We make this distinction because the Schwarzschild criterion for convection $(\nabla>\nabla_a)$ assumes vertical hydrostatic equilibrium, which is not guaranteed in unstable disks.  

\begin{figure}
\plotone{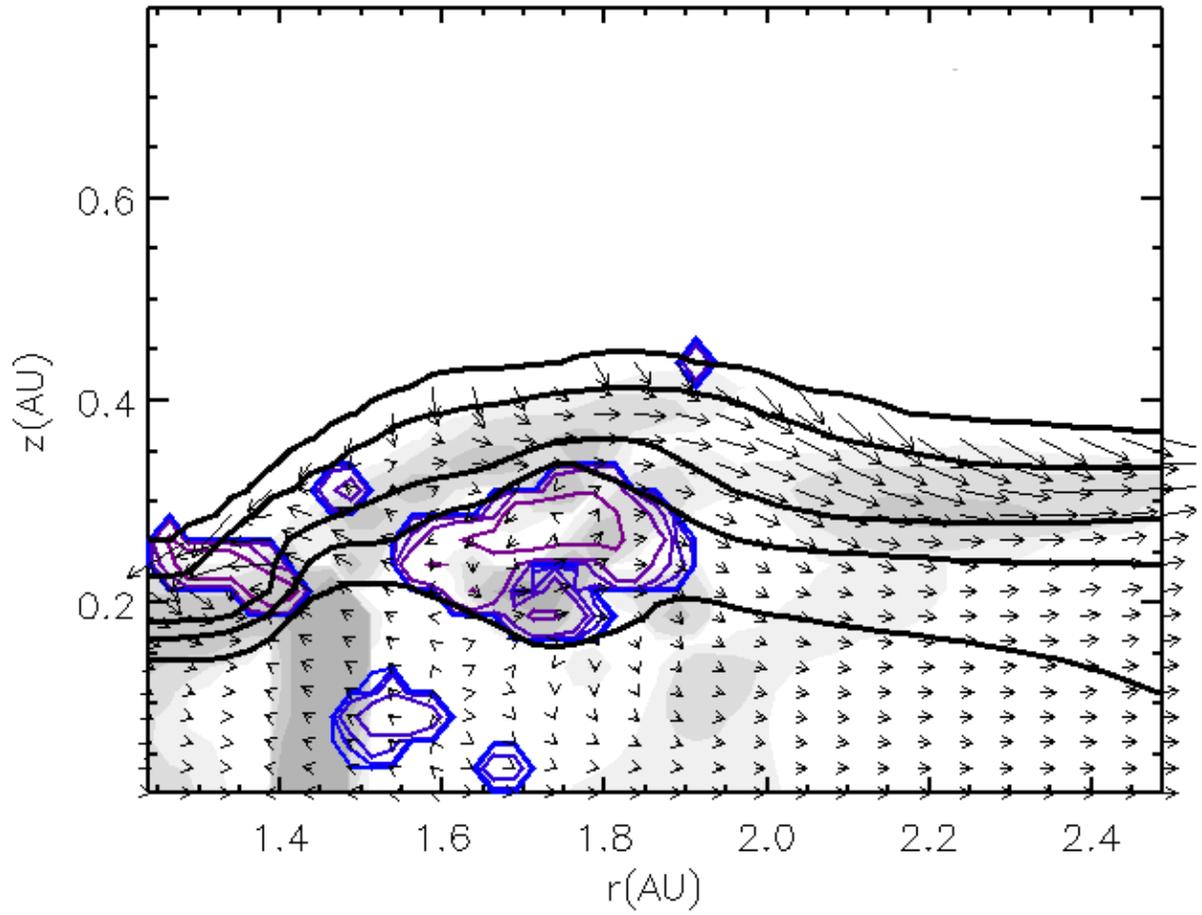}
\caption{Similar to Fig.\ \ref{fig4}, but  at a different location in the high-mass disk with additional contours denoting where the vertical entropy gradient is negative.  }\label{fig18}
\end{figure}

Within some regions of negative entropy gradient, vortical flows are seen, as in Figure \ref{fig18}.  However, shocks are typically found in these regions as well.  Such an alignment suggests that the regions are dynamic, i.e., they are not representative of regions where convection occurs in a stellar interior sense, but are regions that are associated with shock bores and waves.  In addition, the strong shock heating associated with the negative entropy gradients makes it unclear whether such ``dynamic convection'' provides net cooling.

\section{CONCLUSIONS}\label{conclusions}

\subsection{Shock Bores and Waves}

When a strong spiral shock develops in a disk, the shock's highly nonlinear behavior creates a shock bore, where the loss of vertical force balance in the post-shock region results in a rapid vertical expansion of the gas.  The resulting shock structure covers a large range of disk altitudes. The shocks are not limited to the main spiral wave itself.  Jumping gas can fall back onto the pre-spiral shock gas producing breaking waves, vortical flows, and mid- and high-altitude shocks.  The analytic approximations in \S \ref{hstheory} do an accurate job of predicting the height of the disk in the post-shock region as measured in three-dimensional hydrodynamics simulations.

Protoplanetary disks with clumps and spiral arms can have very complex wave dynamics.  As described by several authors \citep[e.g.,][]{bate2003}, waves can transport angular momentum and lead to the formation of gaps.  In addition, as suggested by our simulations, these waves will also provide turbulence in the disk as well as large votical flows, some of which may extend down to the midplane. The combination of radial transport and vortical flows should work to mix the gas and the lighter solids over tenths of an astronomical unit in only an orbit period.   Moreover, the wave dynamics should lead to copious shocks in the disk.

\subsection{Bearing on Reality}

As demonstrated in this paper and in \citet{boley2005proc}, a perturbation must be strong to produce dynamic waves and shock bores in protoplanetary disks.  However, such strong perturbations may be plausible.  The onset of gravitational instabilities could create massive spiral arms and/or clumps that could drive strong spiral shocks throughout the disk, including the inner regions \citep{bossdick}.  Even if the disk, as a whole, is gravitationally stable, there still may be regions in the disk, e.g., a dense ring or annulus formed by a dead zone \citep{gammie1996}, where gravitational instabilities could set in, produce a burst of spiral wave activity, heat the region to stability, and possibly later repeat the process \citep[e.g.,][]{armitage2001}.  Such a transient mechanism could provide strong enough perturbations to drive strong waves and shock bores without opening a gap.  The simulations presented here are not self-consistent inasmuch as they assume the sudden onset of perturbations in disks that are gravitationally stable and are not evolved for a long stretch of time.  These simulations here do, however, demonstrate that shock bores do occur in strong spiral waves, and they provide a basis for comparison with much more complex simulations of GI active disks.  We are currently working on self-consistent simulations to study this phenomenon under more realistic conditions and for longer stretches of time.

\begin{acknowledgments}
A.\ C.\ B.\ was supported in part by an Indiana Space Grant Consortium fellowship and a NASA Graduate Student Research Program fellowship;  R.\ H.\ D.\ was supported in part by NASA grants NAGS-11964 and NNG05GN11G.  The authors would like to thank Don Cox for his helpful suggestions in the preparation of this manuscript.

\end{acknowledgments}

\bibliography{ms}
\bibliographystyle{apj}

\clearpage

\clearpage

\clearpage

\clearpage

\clearpage

\clearpage

\clearpage

\clearpage

\clearpage

\clearpage

\clearpage

\clearpage

\newpage

\clearpage

\clearpage

\clearpage

\clearpage

\clearpage

\clearpage

\begin{table}
\caption{Summary of the parameters for the simulations presented in this study. M$_d$ is the disk mass inside 6 AU, perturbation refers to the type of perturbation used, and $t_{\mathrm{cool}}$ is given in pattern periods.
The last column contains the number of cells used  for each coordinate.  Although two of the calculations have more $z$ cells, the spatial resolution is the same.}
\begin{center}
\begin{tabular}{llllllll}\hline
Disk & Thermal\ Physics\ & M$_d$/M$_{\odot}$ & Perturbation & $r_p$ (AU)& $Q_{\rm min}$& $t_{\mathrm{cool}}$
& ($r$, $\varphi$, $z$) \\\hline
High-mass & Energy eq.\ & 0.143 & $\cos 2\varphi$ & 5.0 & 2 & 4 & (256, 512, 32)\\
&             Isothermal    & 0.143 & $\cos 2\varphi$& 5.0 & 2 & -  & (256, 128, 64)\\
Moderate-mass & Energy eq.\ & 0.0370 & $2\times$2.5 M$_{\rm J}$& 5.2 &  2 & 6& (256, 512, 64)\\\hline
\end{tabular}
\end{center}
\label{models}
\end{table}%

 \end{document}